\documentclass[iop]{emulateapj}
\usepackage{color}

\newcommand{\lya}        {Ly$\alpha$}

\newcommand{\unitcgslum} {erg\,s$^{-1}$}
\newcommand{\unitlco}    {K\,km\,s$^{-1}$\,pc$^2$}
\newcommand{\unitxco}    {$M_{\sun}$ (\unitlco)$^{-1}$}

\newcommand{\kms}        {km\,s$^{-1}$}

\newcommand\bb     {\phn}

\newcommand\co[2]  {CO {\sl J}\,=\,#1--#2}

\newcommand\h      {$^{\rm h}$}
\newcommand\m      {$^{\rm m}$}

\newcommand\coone   {\co{1}{0}}

\newcommand\cofour  {\co{4}{3}}

\newcommand\sst     {SST24 J1434110\,+331733}
\newcommand\um      {\ifmmode\mu{\rm m}\else$\mu${\rm m}\fi}

\newcommand\lpco    {${L^{\prime}}_{\rm\!\!CO}$} 
\newcommand\lfir    {$L_{\rm FIR}$}
\newcommand\msun    {$M_{\sun}$}
\newcommand\lsun    {$L_{\sun}$}
\newcommand\mhtwo   {$M({\rm H_2})$}

\newcommand\perbeam {beam$^{-1}$}
\newcommand\starb   {IRAS12112\,+0305\ }

\slugcomment{Accepted for publication in ApJ.}
\shorttitle{CO Searches in \lya\ Blobs}
\shortauthors{Yang et al.}

\begin{document}

\title{
Constraining Dust and Molecular Gas Properties in L\lowercase{y$\alpha$}
Blobs at \lowercase{$z$} $\sim$ 3 \\
}

\author{
	Yujin Yang\altaffilmark{1},
	Roberto Decarli\altaffilmark{1},
	Helmut Dannerbauer\altaffilmark{2,3},
	Fabian Walter\altaffilmark{1},
	Axel Weiss\altaffilmark{4}, 
	Christian Leipski\altaffilmark{1}, 
	Arjun Dey\altaffilmark{5},
	Scott C. Chapman\altaffilmark{6},
	Emeric Le Floc'h\altaffilmark{2},
	Moire K. M. Prescott\altaffilmark{7},
	Roberto Neri\altaffilmark{8},
	Colin Borys\altaffilmark{9},
	Yuichi Matsuda\altaffilmark{10},
	Toru Yamada\altaffilmark{11},
	Tomoki Hayashino\altaffilmark{12},
	Christian Tapken\altaffilmark{13},
	Karl M. Menten\altaffilmark{4} 
}

% official
\altaffiltext{1}{Max-Planck-Institut f\"ur Astronomie, K\"onigstuhl 17, Heidelberg, Germany}
\altaffiltext{2}{Laboratoire AIM, CEA/DSM-CNRS-Universit\'{e} Paris Diderot, Irfu/Service d'Astrophysique, CEA-Saclay, Orme des Merisiers, 91191 Gif-sur-Yvette Cedex, France}
\altaffiltext{3}{Universit\"at Wien, Institut f\"ur Astronomie, T\"urkenschanz-strasse 17, 1180 Wien, \"Osterreich}
\altaffiltext{4}{Max-Planck-Insitut f\"ur Radioastronomie, Auf dem H\"ugel 69, D-53121 Bonn, Germany}
\altaffiltext{5}{National Optical Astronomy Observatory, 950 N. Cherry Ave., Tucson, AZ 85719}
\altaffiltext{6}{Institute of Astronomy, University of Cambridge, Madingley Road, Cambridge CB3 0HA}
\altaffiltext{7}{Department of Physics, Broida Hall, Mail Code 9530, University of California, Santa Barbara, CA 93106; TABASGO Postdoctoral Fellow}
\altaffiltext{8}{IRAM - Institut de Radio Astronomie Millim{\'e}trique, 300 rue de la Piscine, 38406, Saint-Martin d'H{\`e}res, France}
\altaffiltext{9}{California Institute of Technology, 1200 East California Boulevard, Pasadena, CA 91125, USA}
\altaffiltext{10}{Department of Physics, Durham University, South Road, Durham, DH1 3LE}
\altaffiltext{11}{Astronomical Institute, Tohoku University, Aramaki, Aoba-ku, Sendai, Miyagi 980-8578, Japan}
\altaffiltext{12}{Research Center for Neutrino Science, Graduate School of Science, Tohoku University, Sendai 980-8578, Japan}
\altaffiltext{13}{Leibnitz-Institut f\"ur Astrophysik Potsdam (AIP), An der Sternwarte 16, 14482 Potsdam, Germany}

% short version
%\altaffiltext{1}{Max-Planck-Institut f\"ur Astronomie, Heidelberg, Germany}
%\altaffiltext{2}{Laboratoire AIM, CEA Saclay, France}
%\altaffiltext{3}{Max-Planck-Insitut f\"ur Radioastronomie, Bonn, Germany}
%\altaffiltext{4}{National Optical Astronomy Observatory, Tucson, AZ, USA}
%\altaffiltext{5}{Institute of Astronomy, University of Cambridge, UK}
%\altaffiltext{6}{University of California, Santa Barbara, CA, USA}
%\altaffiltext{7}{Institut de Radio Astronomie Millim{\'e}trique, France}
%\altaffiltext{8}{California Institute of Technology, Pasadena, CA, USA}
%\altaffiltext{9}{Durham University, UK}
%\altaffiltext{10}{Tohoku University, Japan}
%\altaffiltext{11}{Leibnitz-Institut f\"ur Astrophysik Potsdam (AIP), Germany}
%\email{yyang@mpia.de}

\begin{abstract}

In order to constrain the bolometric luminosities, dust properties and
molecular gas content of giant \lya\ nebulae, the so-called Ly$\alpha$
blobs, we have carried out a study of dust continuum and CO line emission
in two well-studied representatives of this population at $z\sim3$: a
\lya\ blob discovered by its strong {\it Spitzer} MIPS 24\um\ detection
(LABd05; Dey et al.~2005) and the Steidel blob 1 (SSA22-LAB01; Steidel
et al.~2000).
We find that the spectral energy distribution of LABd05 is well
described by an AGN-starburst composite template with \lfir\ =
(4.0$\pm$0.5)$\times$$10^{12}$ \lsun, comparable to high-$z$
sub-millimeter galaxies and ultraluminous infrared galaxies.
New Large APEX Bolometer Camera (LABOCA)  870\um\ measurements rule out
the reported Submillimeter Common-User Bolometer Array (SCUBA) detection
of the SSA22-LAB01 ($S_{850\um}$ = 16.8 mJy) at the $>$\,4$\sigma$ level.
Consistent with this, ultra-deep Plateau de Bure Interferometer (PdBI)
observations with $\sim$2\arcsec\ spatial resolution also fail to detect
any 1.2\,mm continuum source down to $\approx$ 0.45\,mJy \perbeam\
(3$\sigma$).  Combined with the existing (sub)mm observations in the
literature, we conclude that the FIR luminosity of SSA22-LAB01 remains
uncertain.
No CO line is detected in either case down to integrated flux limits of
$S_\nu \Delta V$ $\lesssim$ 0.25 -- 1.0 Jy \kms, indicating a modest
molecular gas reservoir, \mhtwo\ $<$ 1--3\,$\times$\,10$^{10}$\msun.
The non-detections exclude, with high significance (12\,$\sigma$),
the previous tentative detection of a \co{4}{3} line in the
SSA22-LAB01.  The increased sensitivity afforded by the Atacama Large
Millimeter/submillimeter Array will be critical in studying molecular
gas and dust in these interesting systems.

\end{abstract}


%----------------------------------------------------------------------
\section{Introduction}

\setcounter{footnote}{0}

\lya\ nebulae (or \lya\ ``blobs'') are extended sources at $z$
$\sim$ 2 -- 6 with typical sizes of the Ly$\alpha$ emission region
of $\gtrsim$\,5\arcsec\ ($\gtrsim$ 50 kpc) and line luminosities
of $L_{\rm{Ly\alpha}}\gtrsim10^{43}$ \unitcgslum, and are some
of the most mysterious astronomical objects \cite[e.g.,][]{Keel99,
Steidel00, Francis01, Matsuda04, Matsuda09, Matsuda11, Dey05, Saito06,
Smith&Jarvis07, Ouchi09, Prescott09, Prescott09th, Yang09, Yang10}.
%%
%% blobs are the 
%% best candidates for the galaxies experiencing interaction with the
%% surrounding IGM at high redshifts.
%%
The nature of the extended \lya\ emission is poorly understood. For
example, \lya\ blobs may represent galaxies forming via cold gas accretion
\citep{Fardal01, Haiman00, Dijkstra&Loeb09, Goerdt10}, galactic-scale
outflows due to star formation \citep{Taniguchi&Shioya00}, or the
result of intense radiative feedback from active galactic nuclei (AGN)
\citep{Haiman&Rees01, Geach09}.

To understand the nature of these \lya\ blobs, we first need to constrain
their energy budget, i.e., the bolometric luminosities of the galaxies
within or in the vicinity of the \lya\ halos and compare the available
energy with the observed \lya\ luminosities.
%%
%% Given that obvious strong UV sources are often not identified in \lya\
%% blobs, infrared (IR) and (sub)mm observations provide complementary
%% information as the energies from embedded sources should be released
%% as the reprocessed thermal dust emission even if they (either star
%% formation or AGNs) are heavily obscured.
%%
Many \lya\ blobs do not appear to contain bright UV continuum sources,
bringing into question the nature of the underlying power source for these
nebulae. Infrared (IR) and (sub)mm observations provide complementary
information in such cases, by detecting and constraining the luminosity
of any dust-enshrouded and obscured power-sources (star-forming or AGN)
within the nebula.
So far, a handful of \lya\ blobs have been detected in the
far-IR (FIR), suggesting that at least some \lya\ blobs contain energy
sources that can power the entire \lya\ luminosity if only a few
percent of their bolometric luminosities are converted to \lya\
radiation \citep{Chapman04,Dey05,Geach05,Geach09,Colbert06,Colbert11}.
On the other hand, it appears that some \lya\ blobs do not contain any
obvious energy sources detectable at mid-IR or millimeter wavelengths
\citep{Nilsson06,Smith08}; in these cases the \lya\ emission has been
attributed to gravitational cooling due to cold mode accretion
\citep{Keres05,Keres09,Dekel09}.

Despite the enormous star formation rates indicated by the FIR
luminosities of some \lya\ blobs (up to $\sim$ 1000\,\msun yr$^{-1}$),
their molecular gas content -- where the stars should form -- has been
mostly unconstrained.
%%
%% Is there enough cold molecular gas to power star formation in
%% the \lya\ blobs?  Or is molecular gas depleted rapidly due to
%% galactic-scale feedback either by the shock driven by superwinds or
%% the photoionization due to AGNs?
%%
Therefore, it is important to understand whether there is enough cold
molecular gas to power star formation in \lya\ blobs or whether the
molecular gas is depleted rapidly due to galactic-scale feedback either
by superwind-driven shocks or photoionization by AGN.
In addition, if detected, carbon monoxide (CO) lines that trace the
molecular gas can be used to probe the kinematics and the excitation
conditions of the surrounding gas.
\lya\ blobs also present an
unique opportunity to investigate the molecular gas content in 
environments that differ significantly from sub-millimeter
galaxies (SMG), high redshift quasars (QSOs) or ultraluminous infrared
galaxies (ULIRGs)
that have been studied in the past \cite[e.g.,][]{Solomon05}.
Most luminous \lya\ blobs share the common property that they reside
in overdense environments \citep{Palunas04, Matsuda04, Prescott08,
Yang09, Yang10, Matsuda09}, suggesting that they may represent the
sites of massive galaxy formation.  Furthermore, the discovery of
spatially-extended \ion{He}{2} $\lambda$1640 emission and very weak
metal emission lines from a \lya\ blob at $z=1.67$ suggests that the
gas in the blob may be of low metallicity \citep{Prescott09}.

In this paper, we present new (sub)mm observations of a $z=2.656$ \lya\
blob discovered by \citet{Dey05} and measure its bolometric luminosity
and dust properties. This blob (\sst; hereafter LABd05) was discovered in
the NOAO Deep and Wide Field Survey Bo\"otes field \citep{Jannuzi&Dey99}
by its strong {\it Spitzer} Multiband Infrared Photometer (MIPS) 24\um\
flux, indicating that as an IR-bright source it may be an ULIRG with a
bolometric luminosity as large as 1--8 $\times$ 10$^{13}$\lsun.
While a 350\um\ flux density of LABd05 had been measured as a part of a
large Caltech Submillimeter Observatory (CSO) Submillimeter High Angular
Resolution Camera (SHARC--II) program \citep{Bussmann09}, the lack of
(sub)mm observations covering both sides of the thermal dust peak of its
spectral energy distribution (SED) prevented a direct constraint on its
bolometric luminosity.

The best-studied \lya\ blob (SSA22-LAB01 at $z=3.09$) in the SSA22
protocluster region discovered by \citet{Steidel00} has been reported
to be associated with a bright SMG from the Submillimeter Common-User
Bolometer Array (SCUBA) observations \cite[e.g.,][]{Chapman04}. However,
follow-up observation with the Submillimeter Array (SMA) at
higher spatial resolution ($\sim$2\arcsec) yielded a non-detection
\citep{Matsuda07}. Furthermore, an Atacama Submillimeter Telescope
Experiment (ASTE) AzTEC single-dish observation with a larger beam
size ($\approx$ 28\arcsec) than SCUBA (15\arcsec) also failed
to detect this source down to $\sim$3 mJy (3$\sigma$) at 1.1\,mm
\citep{Kohno08,Tamura09}.
Thus, the FIR luminosity of SSA22-LAB01 and the exact location of the
dust continuum within the blob is uncertain.  In this paper, we revisit
SSA22-LAB01 with Large APEX Bolometer Camera (LABOCA) measurements
whose wavelength and beam size are close to those of the previous SCUBA
observations.  We also present ultra-deep mm interferometric continuum
observations aiming to detect the possible compact sources within the
blobs and, thus, to pinpoint the exact location of the energy source.

Prior to this work, the only blob targeted with CO observations was
SSA22-LAB01.  \citet{Chapman04} reported a tentative detection of a
CO(4--3) line, which implied a significant molecular gas reservoir,
\mhtwo\ $\sim$ 10$^{11}$\msun.\footnotemark\ 
In this paper, we present a sensitive search for CO emission in this and
the Ly$\alpha$ blob, LABd05.  In \S\ref{sec:observation}, we describe our
dust and CO observations.  In \S\ref{sec:SED_LABd05}, we present the SED
of LABd05 and constrain its dust properties. In \S\ref{sec:SED_LAB01},
we present the single-dish submm observation to verify the previous FIR
measurements, and the deep mm observation of SSA22-LAB01 conducted to
determine the location of its submm continuum emission.  In \S\ref{sec:CO}
and \S\ref{sec:correlation}, we put constraints on the molecular gas
contents of both \lya\ blobs and compare their FIR  and CO luminosities
with those of other high-$z$ galaxies.  In \S\ref{sec:conclusion}, we
summarize the results.  Throughout this paper, we adopt the following
cosmological parameters: $H_0$ = 70\,${\rm km\,s^{-1}\ Mpc^{-1}}$,
$\Omega_{\rm M}=0.3$, and $\Omega_{\Lambda}=0.7$.

\footnotetext{We converted the reported CO flux to the molecular gas
mass using the same assumptions, cosmological parameters adopted in
this paper (Section \ref{sec:CO}) and the CO(4--3)/CO(1--0) brightness
ratio of $\simeq$ 0.5, i.e., the same as \citet{Chapman04}.}

%----------------------------------------------------------------------
\section{Observations and Data Analysis}
\label{sec:observation}

%----------------------------------------------------------------------
%\input{./img/spire_map.tex}
%----------------------------------------------------------------------

%----------------------------------------------------------------------
\begin{figure*}
\epsscale{1.00}
\plotone{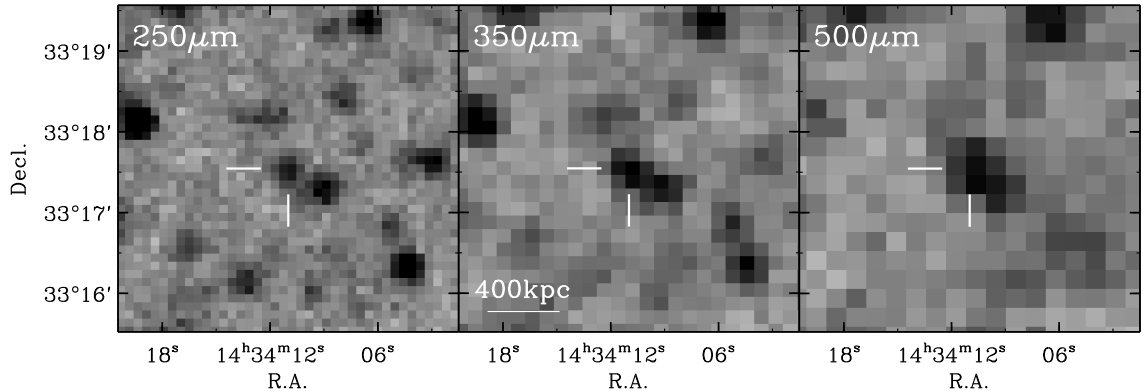}  %% {./img/spire_map.ps} 
\caption{
Herschel/SPIRE 250\um, 350\um, and 500\um\ images of LABd05. The location
of the MIPS 24\um\ source is marked with the bars.
}
\label{fig:spire}
\end{figure*}
%----------------------------------------------------------------------

%----------------------------------------------------------------------

\subsection{Observations of LABd05}

\subsubsection{IRAM-30\,m 1.2\,mm Observation}

We observed LABd05 with Max-Planck Millimeter Bolometer (MAMBO-2), on the
Institut de Radio Astronomie Millim{\'e}trique (IRAM) 30\,m telescope
in 2005 May 18, 19 and 22. MAMBO-2 has a half-power spectral bandwidth
from 210 to 290 GHz, with an effective band center of $\sim$250 GHz
(1.20\,mm) and a beam full-width at half maximum (FWHM) of 10\farcs7.
LABd05 was observed in on-off observing mode under good weather conditions
for $\sim$3\,hr down to an rms of only 0.34\,mJy (1\,$\sigma$) and was
detected with a flux of 2.66\,mJy (i.e., at 7.8\,$\sigma$).

\subsubsection{New Spitzer MIPS 70\um\ and 160\um\ Observations}

\citet{Dey05} reported non-detections of LABd05 at 70\um\ and 160\um\
based on the shallow observations of the Bo\"otes  field carried
out with the MIPS instrument on-board {\it Spitzer}. To further
constrain the photometry of the source at these wavelengths, deeper
pointed observations at 70\um\ and 160\um\ were obtained with the MIPS
``photometry mode'' as part of a General Observer program (ID 20303).
The point spread function sizes (FWHM) are 18\arcsec\ and
40\arcsec\ at 70\um\ and 160\um, respectively.
We reduced these new observations with the MIPS Data Analysis Tool
following the prescriptions described by \citet{Gordon07}, and derived
fluxes for LABd05 using aperture photometry at the position of the
object. However, no detection was obtained even in the deeper data. We
estimated the 3$\sigma$ upper limits as the dispersion of fluxes
measured in similar apertures randomly placed over the background of
each image. Aperture corrections were applied following the MIPS Data
Handbook of the Spitzer Science Center, which led to 3$\sigma$ limits
of 9\,mJy and 51\,mJy at 70\um\ and 160\um, respectively.

\subsubsection{Archival Herschel SPIRE 250\um\ and 350\um\ data}

The Herschel Spectral and Photometric Imaging Receiver (SPIRE) data were
obtained in SPIRE/PACS parallel mode as part of the Herschel Multi-tiered
Extragalactic Survey (HerMES) key project (PI: S.~Oliver). Among
the observations available for the Bo\"otes  field, a total of seven
individual maps cover the source position. The data were processed in a
standard fashion using Herschel Interactive Processing Environment (HIPE)
7.0 and the latest calibration files. While the seven observations were
processed and calibrated individually, the resulting final map was created
by combining all calibrated data and mapping all the scans simultaneously.
We then used a SPIRE source extractor implemented in HIPE \cite[based
on][]{Savage&Oliver07} to find unresolved sources and measure their
fluxes, using data for the approximate shape of the beam from the SPIRE
Observers Manual (version 2.3). Photometric uncertainties were determined
by measuring the pixel-to-pixel rms in an area of 180 arcmin$^2$ around
the science target. This rms was determined in a cleaned map, which had
all the sources detected in the previous step removed.
We show the SPIRE 250, 350, 500\um\ images in Figure \ref{fig:spire}.
The FWHMs of beam profiles are 18\arcsec, 25\arcsec, and 36\arcsec,
respectively.  In the 350\um\ and 500\um\ bands, the LABd05 is blended
with the neighboring galaxy at the south-west direction.  It was not
possible to reliably obtain the 500\um\ photometry due to severe blending.
Our 350\um\ SPIRE photometry (27\,$\pm$\,5 mJy) is consistent with the
previous CSO/SHARC--II observation \cite[$S_{350\um}$ = 37\,$\pm$\,13
mJy;][]{Bussmann09} within the uncertainties. We summarize the photometric
observations in the Table \ref{tab:photometry}.

%----------------------------------------------------------------------
%\subsection{CO Observations for \sst\ and Steidel Blob 1}
\subsubsection{IRAM-30m CO Observations}

LABd05 was observed with the single pixel heterodyne receiver on
the IRAM-30m telescope in UT 2005 May 18 and 21 in good weather
conditions. We used the AB and CD receiver setups, with the AB receivers
tuned to \co{3}{2} (3\,mm band, $\nu_{\rm obs}$ = 94.614 GHz) and the CD
receivers tuned to \co{5}{4} (2\,mm band, $\nu_{\rm obs}$ =
157.675 GHz). The FWHM of beam is 26\arcsec\ and 16\arcsec, respectively.
Average temperatures were 195\,K at 2\,mm and 135\,K at 3\,mm.
Data were taken with a wobbler rate of 0.5 Hz and a wobbler throw of
50\arcsec\ in azimuth. The pointing was checked frequently and was found
to be stable within 3\arcsec\ during the runs. Calibration was done every
12 min with standard hot/cold load absorbers, and we estimate fluxes to
be accurate to within $\sim$10--15\% in both bands.
We used the 512\,$\times$\,1 MHz filter banks for the 3\,mm receiver and
the 256\,$\times$\,4 MHz filter banks for the 2\,mm receivers. As part
of the data reduction, we dropped all scans with distorted baselines,
subtracted linear baselines from the remaining spectra, and then rebinned
to a velocity resolution of 40\,\kms. The total on-source time is
3.4 hr for both CO observations.
The conversion factors from K (${T_{\rm A}}^{\!\!*}$ scale) to Jy at our
observed frequencies are 7.0 Jy\,K$^{-1}$ and 6.1 Jy\,K$^{-1}$ for 2\,mm
and 3\,mm bands, respectively.  The resulting rms noises are 0.4mK (2.8\,mJy)
and 0.45mK (2.8\,mJy) for 2\,mm and 3\,mm bands observations, respectively.
We summarize the CO observations in the Table \ref{tab:CO}.

%%-------------------------------------------------------------------
%% AB: CO(3-2): 3mm 512x1MHz 30M-1M2-B100
%% CD: CO(5-4): 2mm 256x4MHz 30M-4M2-D150
%%-------------------------------------------------------------------

\subsection{Observations of SSA22-LAB01}

\subsubsection{APEX LABOCA 870\um\ Continuum Observations}
\label{sec:LABOCA}

The dust continuum of SSA22-LAB01 was observed at 870\um\ using Large APEX
Bolometer Camera \cite[LABOCA;][]{Siringo09} at the Atacama Pathfinder
Experiment (APEX) in 2011 August 18. Observations were carried out
in the photometric on-off mode under good observing conditions (PWV =
0.5--0.7\,mm). Throw and speed of the chopping secondary were set to
60\,$''$ and 1.5\,Hz, respectively. Chopping was carried out in symmetric
mode with a nodding time of 20\,sec. LABOCA's spatial resolution is
19\farcs2, comparable to the resolution of previous observations with
SCUBA \citep{Chapman04}. Calibration was achieved using Uranus (66.2 Jy
at the time of observations) and the absolute calibration was estimated
to be about 15\%.  Pointing was checked every 40min on the nearby QSO
J2225-045 and found to be stable within 3\farcs6 (rms).  The data were
reduced using the BoA software package. The total integration time on
SSA22-LAB01 is 2h\,20min (on+off, including instrumental overheads). Our
observations yield a non-detection of the source with $S_{870\um}$ =
$-$2.9\,$\pm$\,2.7 ($\pm$\,4.0) mJy \perbeam. The first error is the rms
calculated from error propagation via the rms on the reduced bolometer
time line. The second error in parenthesis is the uncertainty calculated
from the dispersion of the individual scans, therefore represents a more
conservative error estimate.

%----------------------------------------------------------------------
%\input{./img/sed_template_nocomp.tex}
%----------------------------------------------------------------------

%----------------------------------------------------------------------
\begin{figure}
\epsscale{1.0}
\epsscale{1.2}
\plotone{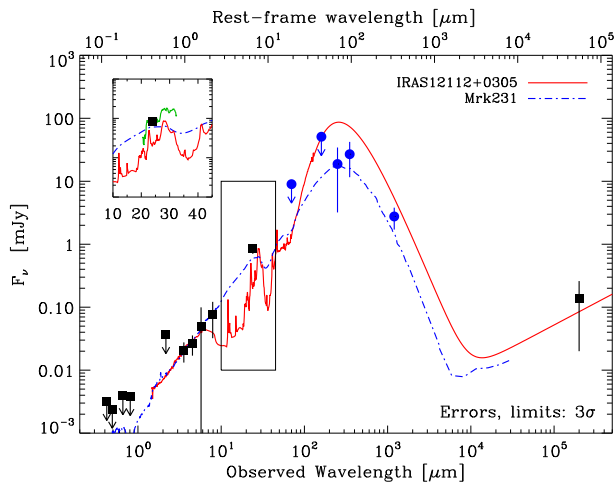}  %% {./img/sed_template_nocomp.ps} 
\caption{
Spectral energy distribution of LABd05 from the rest-frame UV to radio.
The squares represent the photometry compiled by \citet{Dey05} and the
large circles are the {\sl Spitzer} MIPS 70\um, 160\um, Herschel SPIRE
250\um, 350\um, and IRAM-30m/MAMBO-2 1.2\,mm photometry from this work.
The error bars and upper limits are given at 3$\sigma$.
The solid and dot-dashed lines represent the SED templates
of \starb \cite[starburst-dominated;][]{Rieke09} and Mrk 231
\cite[AGN-dominated;][]{Polletta07}. Templates are scaled to match the
3.6\um, 4.5\um, and 5.8\um\ flux densities.  The inset shows the IRS
spectrum from \citet{Colbert11}.
The AGN template shows good agreement with data while the starburst
template (IRAS12112\,+0305) over-estimates the FIR flux by a factor of
$\sim$5.
}
\label{fig:SED_all}
\end{figure}
%----------------------------------------------------------------------

%----------------------------------------------------------------------

\subsubsection{PdBI 1.2\,mm Continuum Observations}

We observed SSA22-LAB01 between May 2010 and January 2011 with the
Plateau de Bure Interferometer (PdBI) in D configuration (Project ID:
T0B2). The on-source observing time, corresponding to the full array of
six antennae, was 5.9~hours. We set the phase center at the reported
SCUBA submm galaxy (SMM J221725.97+001238.9).
%%
%% A radio source is also detected at the J1's location with the 1.4\,GHz
%% flux density of $S_{\rm 1.4\,GHz}$ = 44.4\,$\pm$\,10.1$\mu$Jy
%% \citep{Chapman04}.  The other possible optical counterpart (J2)
%% coincides with an extremely red K band source \citep{Steidel00}
%% and the {\it Spitzer}/IRAC source \citep{Geach07}, LAB1a (R.A.~=
%% 22\h17\m26\fs0, decl.~= 00\degr12\arcmin36\farcs2).
%%
The 1\,mm receiver WideX was tuned at 241.5~GHz, corresponding to 1.25\,mm
in the observed and $\sim$400\,\um\ in the rest-frame. The total bandwidth
of our dual polarization mode observations was 3.6\,GHz. The data were
calibrated through observations of standard bandpass (3C~454.3, 2223-052),
phase/amplitude (2131-021, 2145+067, 2223-052) and flux calibrators
(MWC 349), and reduced with the {\tt GILDAS} software
packages\footnotemark\ {\tt CLIC}
and {\tt MAPPING}. The final map, created using natural weighting,
has an rms noise of 0.15\,mJy integrated over the full bandwidth.
Note that this new observation is approximately four times 
deeper than the previous SMA observations \cite[$\sigma_{\rm rms}$ =
1.4\,mJy at 880\um;][]{Matsuda07}, and one of the deepest observations
ever carried out at $\sim$1\,mm on a single target.  The FWHM of the
beam is 2.4\arcsec$\times$1.6\arcsec\ (18\,kpc$\times$12\,kpc) at
1.25\,mm (P.A.~= 14.9\degr), similar to that of the SMA observations
(2.4\arcsec$\times$1.9\arcsec).

\footnotetext{http://www.iram.fr/IRAMFR/GILDAS}

\subsubsection{PdBI CO Observations}

Observations for SSA22-LAB01 were carried out using the PdBI in May
-- June 2002 (112 GHz, tuned on the CO $J$=4--3 line at $z=3.1$;
Project ID: M021) and in June -- July 2010 (84 GHz, tuned on the CO
$J$=3--2 line. Project ID: U045). Data were collected in compact D
configuration, with baselines ranging between 24 and 113 m.  Phase
calibrators were observed every 20 minutes. The rms phase errors are
$\sim20^{\circ}$. Primary amplitude calibrators were 3C454.3 (variable,
at $\sim 25$ Jy at 3\,mm during our observations), MWC 349 (not variable,
$\sim 1$ Jy) and 2223-052 ($\sim 3.5$ Jy, also used as phase calibrator).
Data at 112 GHz (84 GHz) were collected using the narrow band (WideX)
receiver. System temperatures ranged between 120 and 250 K. Typical
uncertainties in the flux scales and overall calibration are about
10\%. The data processing program used water vapor monitoring receivers
at 22 GHz on each antenna to correct amplitudes and phases for short-term
changes in atmospheric water vapor. We weighted visibilities according to
the inverse square of the system temperature, and natural weights were
applied when creating the maps, in order to boost the sensitivity. The
primary beam size of PdBI is $56''$ ($42''$) at 84 GHz (112 GHz). No
cleaning was applied.  In the case of WideX observation with a large
bandwidth, we also place an upper limit on the continuum ($\sigma_{\rm
rms}$ = 0.15\,mJy (3$\sigma$). We summarize the CO observations in
Table \ref{tab:CO}.

%----------------------------------------------------------------------
%\input{./img/dust_beta.tex}
%----------------------------------------------------------------------

%----------------------------------------------------------------------
\begin{figure}
\epsscale{1.0}
\epsscale{1.2}
\plotone{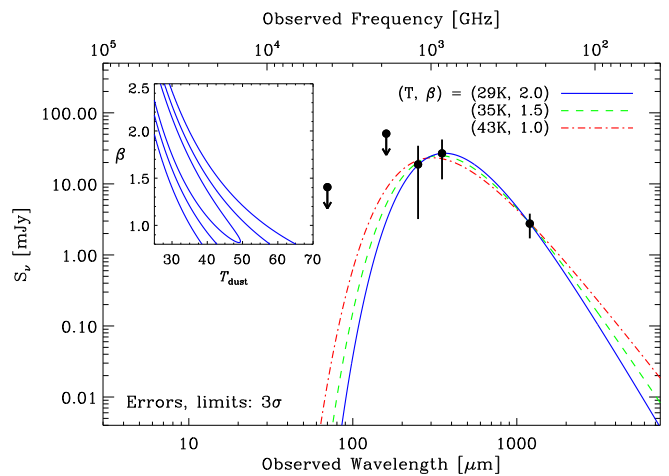} %% {./img/dust_beta.ps}
\caption{
Spectral energy distribution of LABd05 focusing on mid-IR to mm
wavelengths.  New data points from this work are fitted with a modified
blackbody SED with a dust temperature $T_d$ and an emissivity index
$\beta$.  The error bars and upper limits are given at 3$\sigma$.
The left inset shows the likelihood distribution of the fits in $T_d$
and $\beta$ space. The contours represent the 1, 2 and 3 $\sigma$
confidence intervals.
We find a dust temperature, $T_d$ of 44, 35, 29\,K for different $\beta$
= 1, 1.5, and 2, respectively. While there is a degeneracy between $T_d$
and $\beta$, we find that the model with colder dust temperature ($T_d$
$\sim$ 30\,K) provides a better fit to the data.
}
\label{fig:dust_SED}
\end{figure}
%----------------------------------------------------------------------

%----------------------------------------------------------------------

%----------------------------------------------------------------------
\section{Results}

\label{sec:result}

%\subsection{Submm Observation}
\subsection{Spectral Energy Distribution of LABd05}
\label{sec:SED_LABd05}

Using our new FIR photometry from {\it Spitzer} MIPS, Herschel SPIRE and
MAMBO--2, we compare the spectral energy distribution (SED) of LABd05
with those of starbursts and AGNs in order to measure the bolometric
luminosity and investigate whether we can discriminate possible energy
sources of the blob.

Figure \ref{fig:SED_all} shows the SED of LABd05 from optical to
radio wavelength in the observed frame. The squares and large circles
represent the flux measurements from \citet{Dey05} and new constraints
from this study, respectively. We also show the {\it Spitzer}  Infrared
Spectrograph (IRS) spectrum ($\lambda_{\rm obs}$ = 19.5\um -- 38\um)
from \citet{Colbert11}, which reveals that strong polycyclic aromatic
hydrocarbon (PAH) features contribute $\sim$50\% of MIPS 24\um\ flux.
For comparison, we show the SED templates of \starb and the Seyfert 1
galaxy Mrk 231 that are redshifted to $z=2.66$ and scaled to match the
Infrared Array Camera (IRAC) 3.6\um, 4.5\um, and 5.8\um\ flux densities
(rest-frame 1\um--1.5\um).  \starb and Mrk 231 represent local analogs
whose IR SEDs are dominated by star-formation and AGNs, respectively.
We select \starb as a representative template from 11 LIRG/ULIRG
templates compiled by \citet{Rieke09} by searching for a template
that best matches our data.  The Mrk 231 template is also selected by
searching for the best-fit templates from the AGN-dominated templates
compiled by \citet{Polletta07}.

%% As pointed out by \citet{Dey05}, the rest-frame UV fluxes should be
%% regarded as upper limits because these data include the contribution
%% from two galaxies within the blob.

We find that the AGN template (Mrk 231) is able to fit the full SED
of LABd05 from the rest-frame 1\um\ to 350\um\ reasonably well, but
additional PAH emission is required to reproduce the strong 7.7\um\
features in the observed {\it Spitzer} IRS spectrum, and the Mrk 231
template slightly under-predicts the long-wavelength FIR flux densities.
On the other hand, the starburst template overestimates the FIR luminosity
by a factor of $\approx$ 5 and fails to fit the IRAC 8.0\um\ and MIPS
24\um\ data points as discussed by \citet{Dey05}.
Clearly, more sophisticated SED modeling is required and observations
around the rest-frame 2\um--10\um\ are essential for fully constraining
the SED.

Since our new (sub)mm observations cover both sides of the thermal dust
peak, we can constrain the dust temperature and the FIR luminosity of
LABd05. We fit the data with a modified blackbody SED \citep{Hildebrand83}
with the following  functional form at $h\nu/kT_d \lesssim 1$:
\begin{equation}
S_\nu \propto \frac{\nu^{(3+\beta)}}{\exp{({h\nu}/{k T_d})} - 1}
\:,
\end{equation}
where $T_d$ is the effective dust temperature, $\beta$ is the dust
emissivity index with $1\lesssim\beta\lesssim2$.
In Figure \ref{fig:dust_SED}, we show the best-fit models and the
likelihood distribution of $T_d$ and $\beta$ parameters.  While $T_d$
and $\beta$ are degenerate, we find that the model with the higher
$\beta$ and lower dust temperature $T_d$ shows better agreement with
the data.  Insufficient data on the shorter wavelength side of the peak
prevents us from constraining the lower/upper limits on $T_d$/$\beta$,
so we choose $\beta = 2$ as a nominal value and determine the best-fit
dust temperature $T_d$ = $29^{+2}_{-1}$\,K.  For reference, we also show
the fits for fixed $\beta$ = 1 and 1.5 which results in $T_d$ $\simeq$
$44^{+4}_{-3}$ and $35^{+2}_{-2}$\,K, respectively.

Using these SED fits, we derive the far-IR and bolometric luminosity
($L_{\rm FIR}$ and $L_{\rm bol}$) of LABd05.
We find $L_{\rm FIR} (40-1000\um)$ = 
(4.0$\,\pm\,$0.52)\,$\times$10$^{12}$\,$L_{\sun}$ and $L_{\rm bol}$ =
(8.6$\,\pm\,$1.1)\,$\times$10$^{12}$\,$L_{\sun}$ consistent with SMGs
and ultra-luminous infrared galaxies (ULIRGs).
To obtain the bolometric luminosity, we apply the $L_{\rm bol}$ $\approx$
2.15 $L_{\rm FIR}$ correction based on the Mrk 231 SED template.  Note
that the SPIRE 350\um\ flux density, $\nu L_{\nu}$ = (3.5\,$\pm$\,0.6)
$\times$10$^{12}$\,$L_{\sun}$, alone can roughly constrain the $L_{\rm
FIR}$ since the SPIRE band corresponds to the peak of the spectral
energy distribution.
We estimate the dust mass, $M_d$, using
\begin{equation}
M_{\rm dust} = \frac{1}{1+z} 
					\frac{S_{\nu} D_L^2}
					{\kappa_{\nu}^{\rm rest} B(\nu_{\rm rest}, T_d)} \:,
\end{equation}
where $S_{\nu}$ is the observed flux density, $D_L$ is the luminosity
distance, $\kappa_{\nu}^{\rm rest}$ is the rest frame dust mass
absorption coefficient, and $B(\nu, T_d)$ is the Plank function at the
rest frame.  We adopt $T_d$ = 29\,K and $\beta$ = 2 from our best-fit
model and $\kappa_{125\mu m}$ = 26.4 cm$^2$ g$^{-1}$ \citep{Dunne03},
which corresponds to $\kappa_{850\mu m}$ = 0.57 cm$^2$ g$^{-1}$ assuming
$\kappa$ $\propto$ $\nu^{\beta}$.
We find a total dust mass of $M_{\rm dust}$ = (1.4$\,\pm\,$0.7)$\times$10$^9$ $M_{\sun}$.
Note that this dust mass estimate strongly depends on the choice
of $\kappa_\nu^{\rm rest}$ and $T_d$.  The absorption coefficient
$\kappa_{850\um}$ is uncertain to a factor of $\sim$2; e.g.,
$\kappa_{850\um}$ = 0.77 cm$^2$ g$^{-1}$ \citep{Dunne00,Dunne11}
or $\kappa_{850\um}$ = 0.38 cm$^2$ g$^{-1}$ \cite[Milky Way dust
model;][]{Draine03}.  Or if we adopt the lower temperature $T_d$ =
35\,K for $\beta = 1.5$, the dust mass will decrease by a factor of 2.2.

%----------------------------------------------------------------------
%\input{./img/check_SCUBA_det.tex}
%----------------------------------------------------------------------

%----------------------------------------------------------------------
\begin{figure}
\epsscale{1.0}
\epsscale{1.25}
\plotone{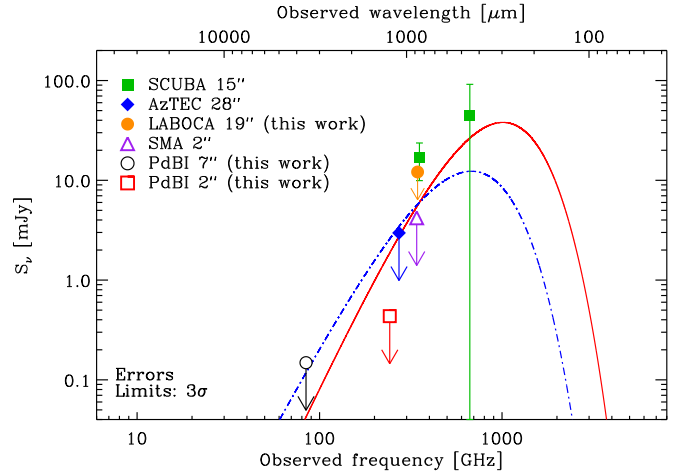} %% {./img/check_SCUBA_det.ps} 
\caption{
FIR flux densities of SSA22-LAB01. Different symbols represent the
(sub)mm observations with different instruments and beams sizes (see
Table \ref{tab:SED_LAB01}).  The filled and open symbols indicate the
flux measurements from single-dish and interferometer observations,
respectively.  All the uncertainties and upper limits are in 3$\sigma$
level.
For illustration purposes, the solid and dot-dashed lines represent
the modified blackbody SEDs normalized at the $S_{850\um}$ = 6\,mJy
with ($T_d$, $\beta$) = (40\,K, 2) and (30\,K, 1.5), respectively. The
upper limits from all other instruments disagree with the SCUBA 850\um\
measurement.
%%
%% reported by \citet{Chapman04}.
%%
}
\label{fig:SED_LAB01}
\end{figure}
%----------------------------------------------------------------------

%----------------------------------------------------------------------

%----------------------------------------------------------------------
%\input{./img/pdbi_1mm.overlay.BV.tex}
%----------------------------------------------------------------------

%----------------------------------------------------------------------
\begin{figure*}
\epsscale{1.1}
\includegraphics[width=0.33\textwidth]{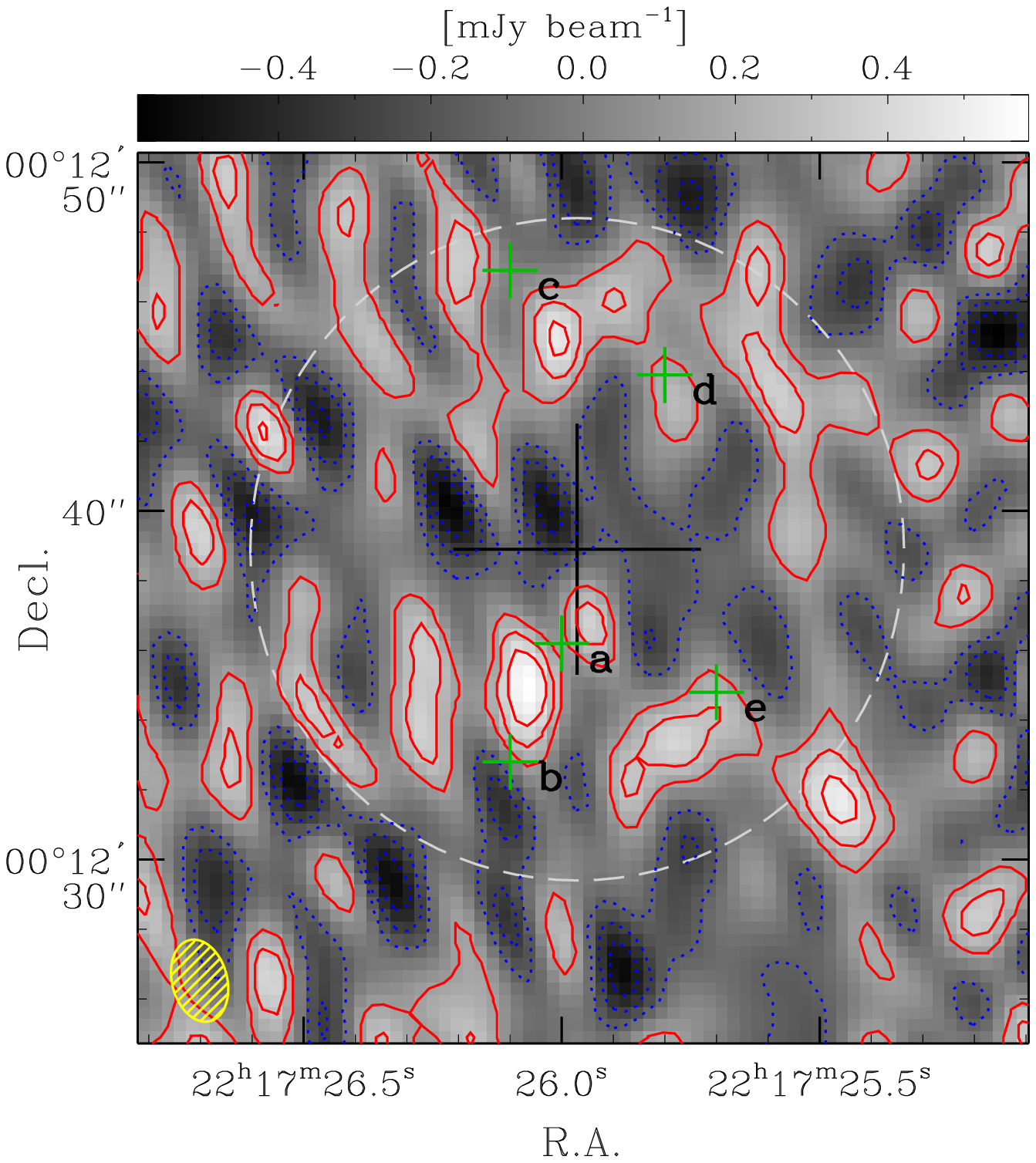} %% {./img/pdbi_1mm.overlay.1mm.ps}
\includegraphics[width=0.33\textwidth]{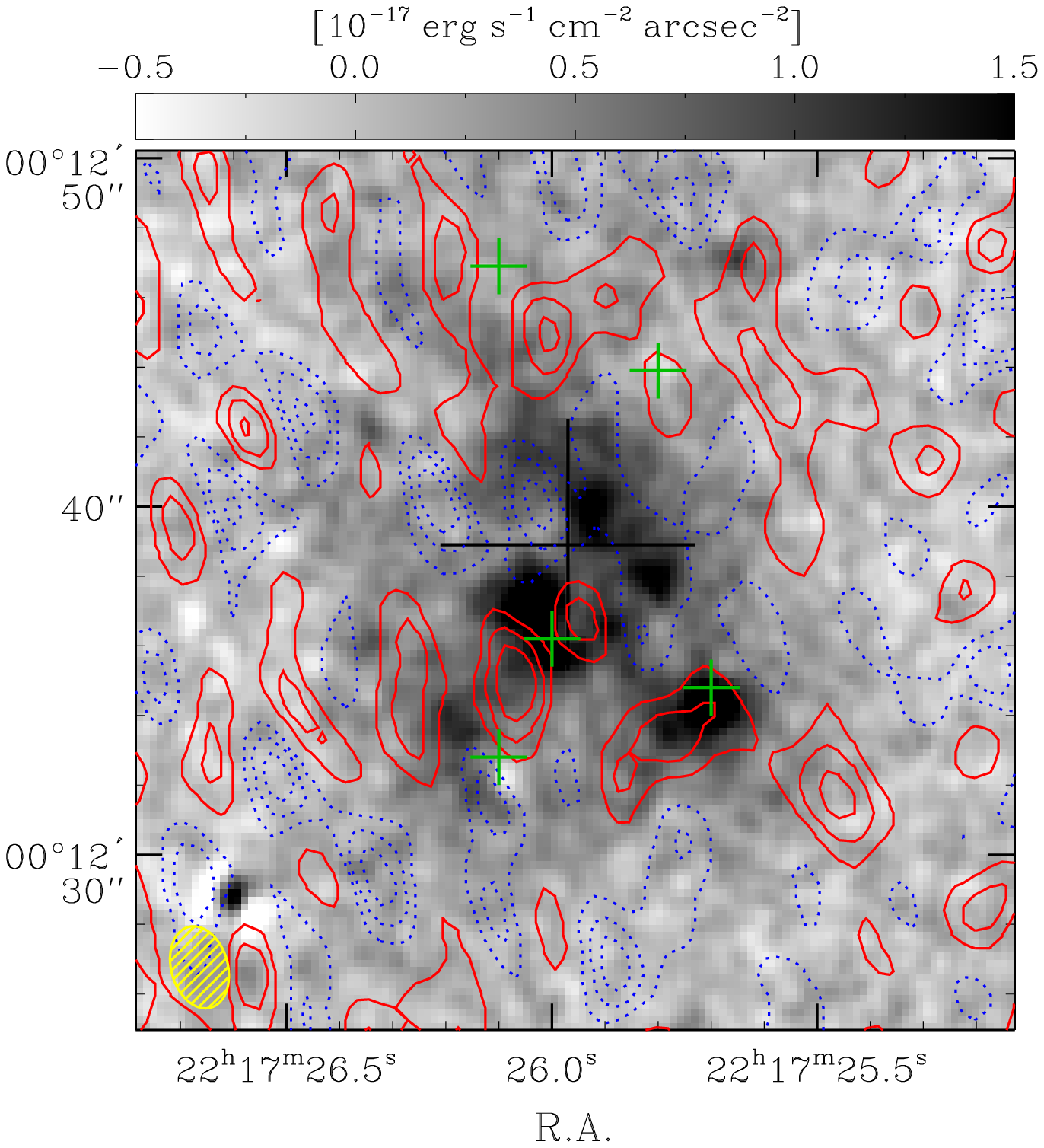} %% {./img/pdbi_1mm.overlay.NB.ps}
\includegraphics[width=0.33\textwidth]{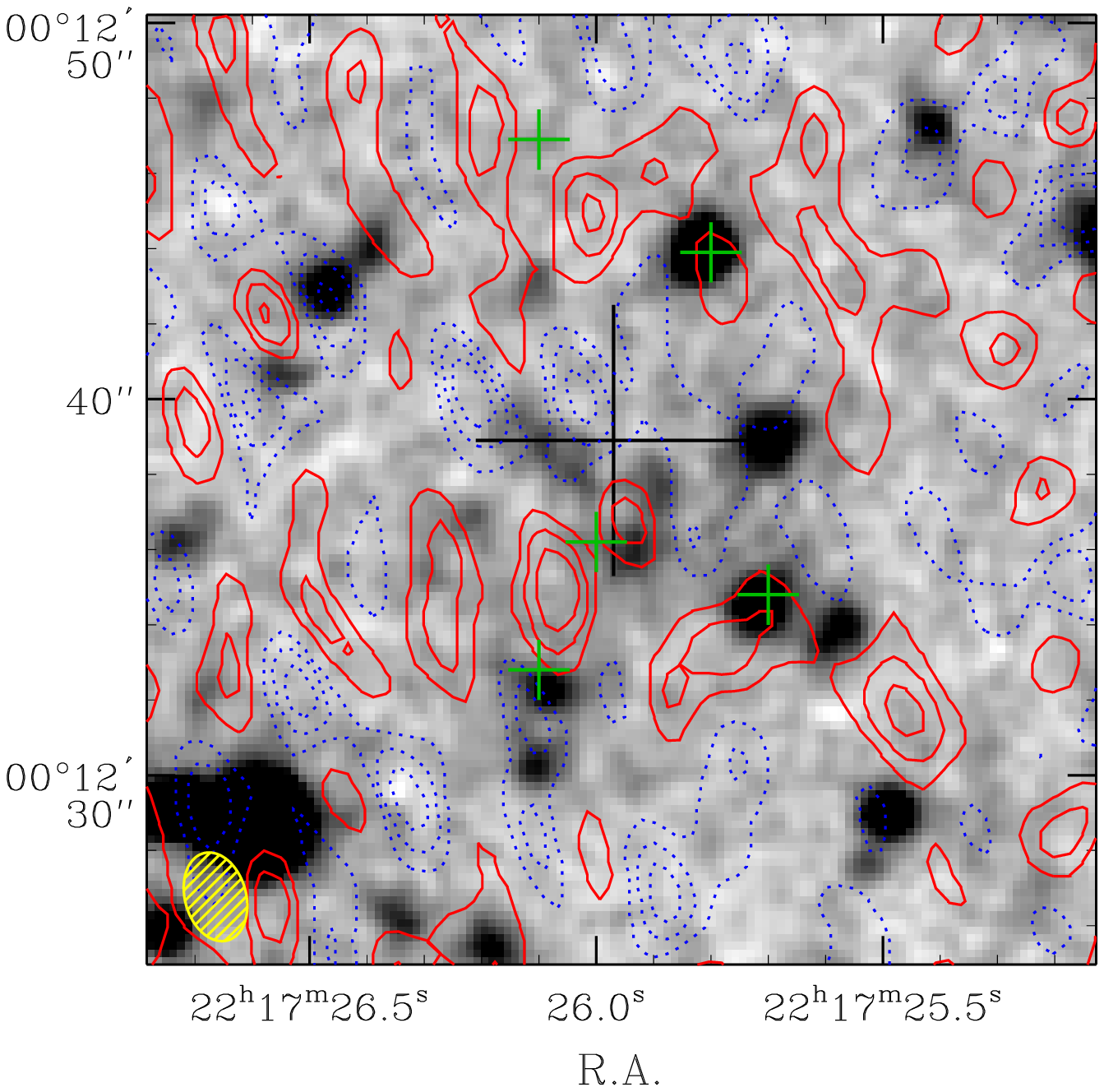} %% {./img/pdbi_1mm.overlay.BV.ps}
\caption{
({\it Left}) 
PdBI 1.2\,mm continuum map of SSA22-LAB01. The synthesized beam has a
FWHM of 2\farcs4$\times$1\farcs6 (P.A.=14.9\degr). The phase center
of the PdBI is marked with a large cross --- its size indicates the
positional uncertainty of the LABOCA pointing.  The LABOCA beam (FWHM =
19\arcsec) is represented by the dashed circle.  The small crosses represent
the locations of the {\it Spitzer} IRAC sources \citep{Geach07}. No
significant continuum source is detected above 3$\sigma$ level at the
location of these galaxies.
({\it Middle and right}) 
PdBI 1.2\,mm continuum contours superimposed on the Subaru/Suprime-Cam
\lya\ and {\sl BV} broadband images \citep{Matsuda04}, respectively.
The contours show the $-$3, $-$2, $-$1, 1, 2, 3$\sigma$ with $\sigma$
= 0.15 mJy beam$^{-1}$.
}
\label{fig:continuum_LAB01}
\end{figure*}
%----------------------------------------------------------------------

%----------------------------------------------------------------------

\subsection{Non-detection of Continuum Emission in SSA22-LAB01}
\label{sec:SED_LAB01}

SSA22-LAB01 was not detected down to $S_{870\um}$ = 8.1 -- 12\,mJy
\perbeam\ ($3\sigma$) with new APEX/LABOCA single-dish observations
which have similar beam size (19\arcsec) to the previous SCUBA 850\um\
observations (15\arcsec).  \citet{Chapman04} reported that SSA22-LAB01
contains a bright submm galaxy: SMM J221725.97+001238.9 with the flux
densities of $S_{850\um}$ = 16.8\,$\pm$\,2.9\,mJy and $S_{450\um}$ =
45.1\,$\pm$\,15.1\,mJy, respectively.\footnotemark\ \ 
If we assume simple model SEDs with ($T_d$, $\beta$) = (40\,K, 2) and
(30\,K, 1.5), we expect an 870\um\ flux density of $S_{870\um}$ = 15.7
-- 16.1 mJy, which would be detected by LABOCA with 4\,--\,6$\sigma$
significance depending on the LABOCA error estimates (\S\ref{sec:LABOCA}).
%%
%% If we consider the uncertainties of the SCUBA observations, the
%% LABOCA observations rule out these SCUBA measurements with 3.2$\sigma$
%% confidence level.
%%
This non-detection is consistent with essentially all measurements
of the source in the literature, in particular the SMA observations by
\citet{Matsuda07}.  In that study, SSA22-LAB01 was also not detected down
to $S_{880\um}$ = 4.2\,mJy \perbeam\ (3$\sigma$) with a $\sim$2\arcsec\
spatial resolution. They proposed, as explanation, that the dust emission
(reported by SCUBA) is extended on spatial scales larger than 4\arcsec\
such that the interferometer observation would resolve out this smooth
component.  However, the original SCUBA observation is not verified by
our LABOCA observation, for which this argument does not hold.
Furthermore, an ASTE-AzTEC single-dish observation with a larger beam
size (FWHM $\approx$ 28\arcsec) than SCUBA (15\arcsec) also failed
to detect dust continuum down to $\sim$3 mJy (3$\sigma$) at 1.1\,mm
\citep{Kohno08,Tamura09}, thus contradicting the SCUBA measurement at the 
4.5$\sigma$ confidence level.
In Figure \ref{fig:SED_LAB01}, we show the upper limits derived from our
LABOCA and PdBI observations with the existing (sub)mm observations
of the SSA22-LAB01.  These observations are summarized in Table
\ref{tab:SED_LAB01}.
The solid and dot-dashed lines represent the two model SEDs normalized at
$S_{850\um}$ = 6\,mJy, showing the possible range of the extrapolated flux
densities at various wavelength.  The upper limits from all other (sub)mm
observations (both single-dish and interferometer) are inconsistent with
the SCUBA measurements.  Therefore, we conclude that the previous SCUBA
measurements are not reliable.

\footnotetext{Slightly different values for the SCUBA 850\um\ and 450\um\
fluxes are reported in \citet{Chapman01, Chapman04, Chapman05}. Since
all the flux measurements agree with each other within uncertainties,
we adopt the values reported by \citet{Chapman04} throughout the paper.}

Given that SSA22-LAB01 contains multiple galaxies and that the total star
formation rate within the blob is at least $\sim$220\,\msun yr$^{-1}$
\citep{Matsuda07}, it is critical to identify galaxies by their
dust emission within the \lya\ halo which produce most of bolometric
luminosity.
In order to detect dust emission from individual galaxies within 
SSA22-LAB01, we compare our deep PdBI 1.2\,mm continuum map with
the locations of the UV and {\it Spitzer} IRAC sources identified by
\citet{Chapman04} and \citet{Geach07}.
In Figure \ref{fig:continuum_LAB01}, we show the PdBI 1.2\,mm continuum
map of SSA22-LAB01 and overlay the contours of this map on the \lya\ and
{\sl BV} broadband images \citep{Matsuda04}. The pointing center of the
LABOCA observation is marked with a large cross (positional uncertainty
of 3\farcs6) and a dashed circle gives the beam size of 19\arcsec.  This
pointing center is the same as the phase center of the PdBI observations.
We show locations of the {\it Spitzer}/IRAC sources marked as LAB
1\,$abcde$ \citep{Geach07} in Figure \ref{fig:continuum_LAB01}.  Out of
these five IRAC sources, three sources (LAB 1\,$a$, $b$, and $e$) are
likely associated with the SSA22-LAB01 \citep{Geach07,Uchimoto08}.
We find that no significant emission is detected from these three galaxies
while there is a $\sim$4$\sigma$ peak (R.A.~= 22\h17\m26\fs06, decl.~=
00\degr12\arcmin35\farcs1) between LAB 1$a$ and 1$b$. Furthermore, none of
$\sim$3$\sigma$ peaks within the FOV of our 1.2\,mm map agrees with other
UV sources in the field (right panel in Figure~\ref{fig:continuum_LAB01}).
Therefore, we conclude that none of the known optical or near-IR emitting
galaxies within the SSA22-LAB01 \lya\ halo are detected at 1.2\,mm down
to 0.45 mJy \perbeam\ (3$\sigma$). Future observations with the Atacama
Large Millimeter/submillimeter Array (ALMA) will be required to verify
this tentative $\sim$4$\sigma$ peak.

Without any formal detection at (sub)mm wavelengths, the FIR luminosity
of SSA22-LAB01 remains highly uncertain.  We place upper limits on
the FIR luminosity of SSA22-LAB01 using simple model SEDs to provide
the possible range of \lfir.  If we normalize the SEDs at the 870\um\
3$\sigma$ upper limit, we obtain \lfir\ $<$ 3.1$\times$10$^{12}$\,\lsun\
and 1.2$\times$10$^{13}$\,\lsun\ depending on the choice of the model SED,
($T_d$, $\beta$) = (30\,K, 1.5) and (40\,K, 2), respectively.  If we
adopt our 1.2\,mm PdBI measurement, the individual galaxies within the
blob have \lfir\ $<$ (0.39 -- 2.1) $\times$10$^{12}$ \lsun\ (3$\sigma$).

%----------------------------------------------------------------------
%\input{./img/ssa22-co.tex}
%----------------------------------------------------------------------

%----------------------------------------------------------------------
\begin{figure*}
\epsscale{1.20}
\epsscale{0.85}
\plotone{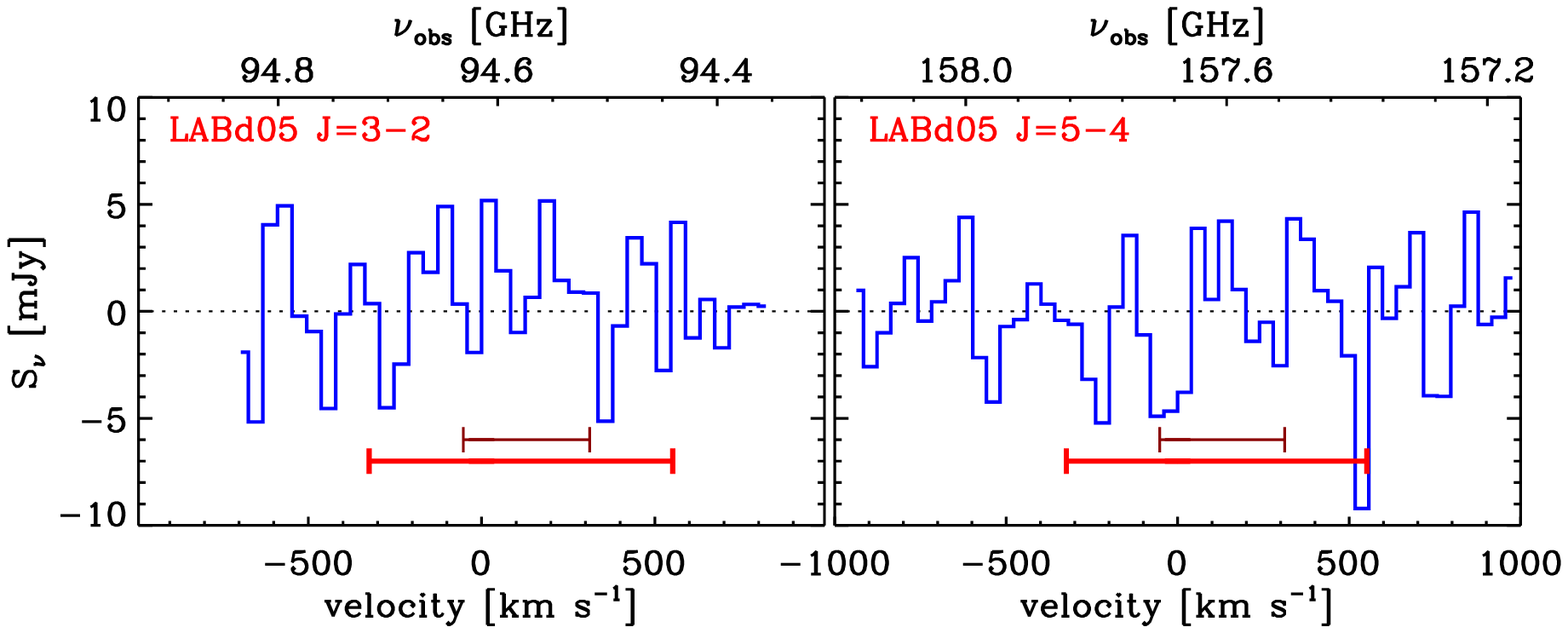}\\[1.0ex]  %% {./img/sst24-co.ps}\\[1.0ex]
\plotone{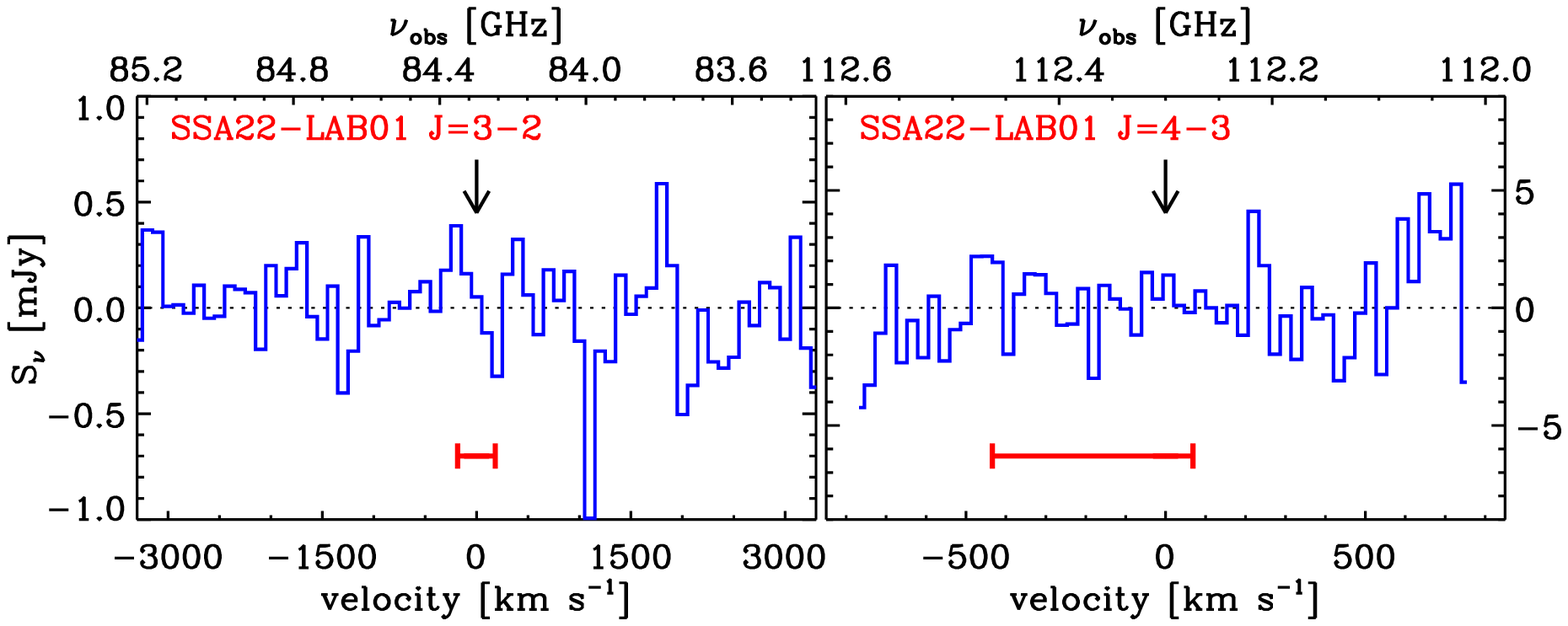}           %% {./img/ssa22-co.ps} 
\caption{
({\it Top}) IRAM-30m CO spectra for \co{3}{2} and \co{5}{4} transitions
for LABd05. Both spectra have velocity resolutions of $\delta v$ =
40\,\kms. The thick and thin horizontal bars represent the expected
velocity range estimated from the \lya\ and \ion{He}{2} emission line
in the optical \citep{Dey05}.
({\it Bottom}) PdBI CO spectra for \co{3}{2} and \co{4}{3} transitions
of SSA22-LAB01 with $\delta v$ = 100\,\kms\ and 27\,\kms, respectively.
Note that the scales of $y$-axis are different between two panels. 
The thick horizontal bars represent the expected velocity range estimated
from the \lya\ line \citep{Bower04}. The vertical arrows indicate the
location of the previous tentative CO detection \citep{Chapman04}.
No CO line is detected in either case.
}
\label{fig:CO_spectra}
\end{figure*}
%----------------------------------------------------------------------

In addition to the PdBI 1.2\,mm continuum observation, we also place
an upper limit on the 3\,mm flux density using the PdBI CO observation
(see Section \ref{sec:CO} and Table \ref{tab:SED_LAB01}). No continuum
source is detected in the integrated channel map over 4\,GHz bandwidth,
resulting in 3$\sigma$ limit of 0.15 mJy \perbeam\ with a beam size of
8\farcs4 $\times$ 5\farcs4.

\subsection{Molecular Gas Content of LABd05 and SSA22-LAB01}
\label{sec:CO}

%----------------------------------------------------------------------

In Figure \ref{fig:CO_spectra}, we show the spectra of the CO lines
for the LABd05 and the SSA22-LAB01 obtained from the IRAM-30m and PdBI,
respectively.
For LABd05, we show the spectra for \co{3}{2} and \co{5}{4} lines with
a velocity resolution of $\delta v$ = 40\,\kms. For SSA22-LAB01,
we show \co{3}{2} and \co{4}{3} transitions with $\delta v$ =
100\,\kms\ and 27\,\kms, respectively.
The rms noise per channel ranges from 0.5 to 2.8 mJy \perbeam\  
(Table \ref{tab:CO}) depending on the instruments and the transitions.

Because of the width of spectral bandwidth ($\sim$2000\,\kms), knowing
the accurate redshift is critical in this CO detection experiment.
In Figure \ref{fig:CO_spectra}, we show the velocity range of CO lines
that corresponds to the range spanned by \lya: $z$ = 2.656 $\pm$ 0.006
for LABd05 \citep{Dey05} and $z$ = 3.102 $\pm$ 0.005 for SSA22-LAB01
\citep{Bower04,Matsuda05}.
Given that \lya\ emission lines from star-forming galaxies at $z=2-3$
could be redshifted against the optically thin nebular lines (i.e.,
systemic velocity) by 250 -- 1000\,\kms\ \cite[e.g.,][]{Pettini01,
Steidel04, Steidel10} due to either galactic-scale outflows or absorption
by the intervening intergalactic medium, we note that CO lines, which are
also expected to emit near the systemic velocity \cite[e.g.,][]{Greve05},
could be located blueward (higher $\nu_{\rm obs}$) of the horizontal bars.
For LABd05, we also show the velocity range corresponding to the optically
thin \ion{He}{2} $\lambda$1640 emission that originates from the central
region of the \lya\ nebula.
At least for LABd05, these two estimates for CO line centers are in good
agreement, therefore we conclude that it is unlikely that any potential
CO line would fall outside of the spectral bandwidth covered by our
IRAM-30m observations.
We note that, for SSA22-LAB01, the \cofour\ spectra obtained with the old
narrow--band receivers at PdBI were taken with a limited bandwidth and
no other optically thin emission lines were available to double-check
the systemic redshift.  On the other hand, the PdBI \co{3}{2} spectrum
obtained with the new WideX correlator has a very wide spectral coverage
(4\,GHz; 14000\,\kms), enough to cover any plausible velocity offsets
between \lya\ and CO lines.  We conclude that no significant CO emission
is detected above the rms noise near the expected velocities in both
systems.

\begin{figure}
\epsscale{0.95}
\epsscale{1.15}
\plotone{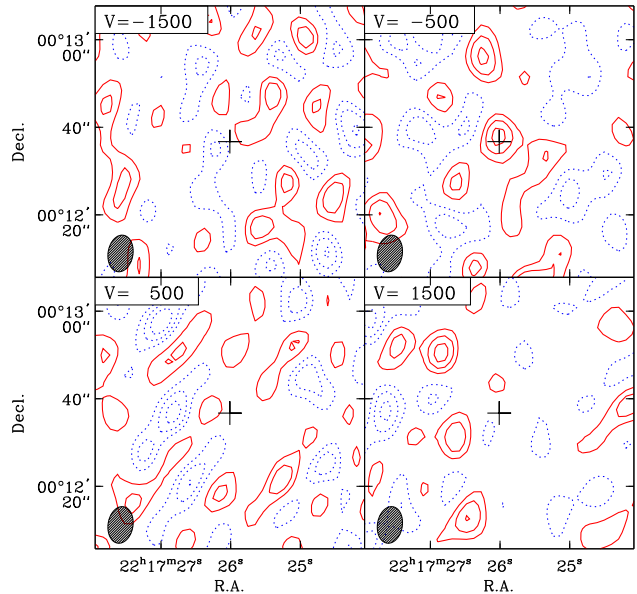} %% {./img/channel_map_1000kms.ps} 
\caption{
Integrated channel map of the \co{3}{2} line for SSA22-LAB01. Each channel
map is integrated over a 1000\,\kms\ bandwidth centered at the velocities
shown in the upper left corners.  The contours are at $-$3, $-$2, $-$1,
1, 2, and 3$\sigma_{\rm rms}$, where $\sigma_{\rm rms}$ is the rms noise
per 1000\kms\ velocity width (0.172 mJy beam$^{-1}$).  The phase center is
marked with a cross in each panel.  While there is a possible 3$\sigma$
detection near the phase center at the $v$ = $-$500\,\kms\ channel, no
significant CO line is detected at above 3$\sigma$ level per 100\,\kms\
channel in that bin.
}
\label{fig:CO_channel_LAB01}
\end{figure}

In the case of SSA22-LAB01, we also inspect whether any tentative
source can be detected in the integrated channel maps.  In Figure
\ref{fig:CO_channel_LAB01}, we show the channel maps summed over $\Delta
V$ = 1000\,\kms\ intervals centered at $-$1500, $-$500, 500, 1500\,\kms\
from the \lya\ velocity center.  While there could be a possible 3$\sigma$
detection near the phase center in the $v$ = $-$500\,\kms\ channel map
(upper right panel of Figure \ref{fig:CO_channel_LAB01}), this marginal
detection in the integrated map corresponds to only $\sim$1.7$\sigma$
signal per 100\,\kms\ channels.  Because we are not able to extract
a reliable spectrum from this tentative source, we conclude that no
significant CO line is detected above the 3 $\times$ rms noise near the
expected velocities.  More sensitive observations using ALMA will test
whether or not this tentative emission is real.

Our new PdBI observation for SSA22-LAB01 rules out the previous tentative
CO detection at high significance.  \citet{Chapman04} reported a tentative
(3.2\,$\sigma$) detection of a \co{4}{3} line from SSA22-LAB01
using the Owens Valley Radio Observatory Millimeter Array. The reported
intensity is $S_{\nu}$ $\sim$ 10 mJy at the peak with a line width of
$\sim$400\,\kms.
In Figure \ref{fig:CO_spectra}, we indicate the location of earlier
tentative CO detection with arrows.  As described in detail below, our
3$\sigma$ limit for the integrated \co{4}{3} flux from PdBI is $S_{\nu}$
$\Delta V$ $<$ 0.62 Jy \kms, i.e., we exclude the previous detection at
high significance (12$\sigma$).

Using these non-detections, we put constraints on CO line luminosities
and molecular gas mass of the blobs. 
Following \citet{Solomon05}, the CO luminosity $L^\prime_{\rm CO}$
(in \unitlco) is given by 
%----------------
\begin{equation}
L^\prime_{\rm CO} = 	3.25 \times 10^7 \: 
S_{\rm CO} \: \Delta V \: \nu^{-2}_{\rm obs}\:  D_L^2 \: (1+z)^{-3} \:\: ,
\label{eq:lpco}
\end{equation}
%----------------
where $\nu_{\rm obs}$ is the observing frequency (in GHz), $D_L$ is the
luminosity distance (in Mpc) to the source at a redshift $z$.
For our non-detection, we adopt a $3\sigma$ upper limit on the
velocity-integrated flux $S_{\rm CO} \Delta V$ $\equiv$
3$(\Delta V/\delta v)^{1/2}$ ($\delta v\:\sigma_{\rm rms}$) in Jy\,\kms, 
where $\Delta V$ is the CO line width, where $\delta v$ is the channel
bandwidth, and $\sigma_{\rm rms}$ is the rms noise value per channel,
respectively.
As the CO line width is unknown, we adopt $\Delta V$ = 400\,\kms, 
consistent with FWHM = 300 -- 450\,\kms\ measured from the optically thin
H$\alpha$ lines in two other \lya\ blobs in the Extended Chandra Deep
Field South \citep{Yang11}.
For LABd05, we obtain $3\sigma$ CO line luminosity limits, 
${L^\prime}_{\rm\!\!CO(5-4)}$ $<$ 1.37$\,\times\,$$10^{10}$\,$(\Delta V/400)^{1/2}$ \unitlco\ 
and 
${L^\prime}_{\rm\!\!CO(3-2)}$ $<$ 3.90$\,\times\,$$10^{10}$\,$(\Delta V/400)^{1/2}$ \unitlco.
In the case of the SSA22-LAB01, we obtain the upper limits of
${L^\prime}_{\rm\!\!CO(4-3)}$ $<$ 1.6$\,\times\,$$10^{10}$\,$(\Delta V/400)^{1/2}$ and 
${L^\prime}_{\rm\!\!CO(3-2)}$ $<$ 1.5$\,\times\,$$10^{10}$\,$(\Delta V/400)^{1/2}$ \unitlco.

%for {\sl J}=4--3 and {\sl J}=3--2 transitions, respectively.

%----------------------------------------------------------------------
%\input{./img/channel_map_1000kms.tex}
%----------------------------------------------------------------------

%----------------------------------------------------------------------
%----------------------------------------------------------------------

%----------------------------------------------------------------------

The upper limits for the total molecular gas mass, $M({\rm H_2})$, can
be derived using a CO--to--H$_2$ conversion factor between \co{1}{0}
luminosity and molecular gas mass, $X_{\rm CO}$.  Here we adopt $X_{\rm
CO}$ $\approx$ 0.8 $M_{\sun}$ (\unitlco)$^{-1}$, which was proposed as
appropriate for starburst environments found in ultra-luminous infrared
galaxies \citep{Downes&Solomon98}.
If we assume a constant brightness temperature for the different CO
transitions, i.e., \coone\ luminosity as defined by Eq.~(\ref{eq:lpco}) is
the same as those of upper ({\sl J+1}$)\rightarrow\,${\sl J} transitions,
the upper limits on the molecular gas mass can be given as
$M({\rm H_2})$ $<$ 3.1 and 1.2$\,\times\,$ 
$10^{10} M_{\sun} (\Delta V/400)^{1/2} (X_{\rm CO}/0.8)$ 
for LABd05 and SSA22-LAB01, respectively.
Note that, for submm galaxies and QSOs in a similar redshift range,
the gas has been found to be thermalized at least up to {\sl J}=3--2
transition as the ratio 
${L^\prime}_{\rm\!\! CO(3-2)}$/${L^\prime}_{\rm\!\! CO(1-0)}$ is close to
unity ($\approx 0.8$) \cite[e.g., ][]{Weiss07}.  Therefore, we use only
the {\sl J}=3--2 transition to estimate limits on ${L^\prime}_{\rm\!\!
CO(1-0)}$ and \mhtwo.
We note that our final \mhtwo\ estimates are strongly dependent not only
on the intrinsic excitation, but also on the choice of $X_{\rm CO}$. For
example, if we adopted $X_{\rm CO}$ $\approx$ 4.5 \unitxco, the Milky
Way value, the resulting molecular gas mass limits would increase by a
factor of $\sim$5.

%% In order to estimate the \coone\ luminosity from the higher {\sl J}
%% transitions, we use the CO line SED \citep{???} which is the plot
%% of CO peak flux density ($S_{\rm CO}$) versus the rotational quantum
%% numbers $J_{\rm upper}$.  Figure \ref{fig:CO_SED} shows the 3$\sigma$
%% upper limits on the CO peak flux density in two transitions compared
%% to the possible CO SEDs derived from models.

\subsection{\lfir--\lpco\ Correlation}
\label{sec:correlation}

With the FIR luminosity and the upper limit on CO luminosity in hand, we
investigate whether \lpco\ and $L_{\rm FIR}$ of the LABd05 are consistent
with the known $L_{\rm FIR}$--\lpco\ scaling relations.  For SSA22-LAB01,
the ratio between \lfir\ and \lpco\ remains unconstrained so far.
For nearby starburst and spiral galaxies with \lfir\ $\lesssim$
10$^{12}$\lsun, the correlation has a form,
$\log L_{\rm FIR}$ = ($1.26 \pm 0.08$) $\times$ $\log L^{\prime}_{\rm CO}$ $-$ 0.81 
\citep{Gao&Solomon04}. 
When ULIRGs, SMGs, radio galaxies (RGs), and QSOs are included, 
the relation becomes slightly steeper,
$\log L_{\rm FIR}$ = ($1.39 \pm 0.05$) $\times$ $\log L^{\prime}_{\rm CO}$ $-$ 1.76, 
but still holds over 5 orders of magnitude in $L_{\rm FIR}$ \citep{Riechers06}.
For LABd05 with $L_{\rm FIR}$ =
(4.0\,$\pm$\,0.5)\,$\times$\,$10^{12}$ $L_{\sun}$, we expect \lpco\ =
(2.2\,$\pm$\,1.9)\,$\times$\,10$^{10}$\,\unitlco\ for the ULIRG/SMG/RG/QSO
relation.  Therefore, we find that our 3$\sigma$ upper limit on \lpco\ for
LABd05 is consistent with the known \lfir--\lpco\ scaling relations.  In
terms of the continuum-to-line luminosity ratio, we find that \lfir/\lpco\
$>$ 100, which is consistent with the \lfir/\lpco\ = 125\,--\,240 \lsun
(\unitlco)$^{-1}$ for the galaxies with \lfir\ $\sim$ 10$^{12-13}$\lsun.
Note that due to the non-linear relation between \lfir\ and \lpco,
the ratio \lfir/\lpco\ has a large spread depending on \lfir.

\clearpage
%----------------------------------------------------------------------
\section{Summary and Concluding Remarks}
\label{sec:conclusion}

We have obtained IR and (sub)mm observations of the two best-studied \lya\
blobs, LABd05 and SSA22-LAB01, in order to constrain their energy budgets
(i.e., the bolometric luminosity) and dust properties.
We find that LABd05 has a high FIR luminosity, \lfir = 4.0 $\times$
10$^{12}$\lsun, comparable to values found for high-$z$ SMGs and
ULIRGs. The NIR-to-FIR SED of LABd05 is well described by the
AGN-starburst composite template (Mrk 231).
For SSA22-LAB01, no 870\um\ continuum is detected down to 8.1--12\,mJy
\perbeam\ (3$\sigma$) in the LABOCA single-dish observation contrary to
the originally reported SCUBA measurement.  To detect the dust emission
from individual galaxies within SSA22-LAB01, we obtained a very deep
1.2\,mm observation of SSA22-LAB01. No 1.2\,mm continuum is detected down
to $\approx$ 0.45 mJy \perbeam\ (3$\sigma$) at $\sim$2\arcsec\ resolution.
Combined with the existing extensive (sub)mm observations from
literature, we conclude that the previously published SCUBA detection
\citep{Chapman04} is not reliable.  We place 3$\sigma$ upper limits
of \lfir\ $<$ (3.1 -- 12) $\times$ 10$^{12}$ \lsun\ for the total FIR
luminosity of the blob, and \lfir\ $<$ (0.39 -- 2.1) $\times$ 10$^{12}$
\lsun\ for the individual galaxies within the blob.

To investigate the molecular gas content in \lya\ blobs, we carried out
a sensitive search for CO lines in the two \lya\ blobs.  No CO line is
detected down to an integrated flux limit of $S_\nu \Delta V$ $\lesssim$
0.25 -- 1.0 Jy\,\kms\ constraining the molecular mass to be less than
\mhtwo\ $\approx$ (1--3)\,$\times$\,10$^{10}$\msun\ (3$\sigma$ limit),
assuming a constant brightness temperature and a CO--to--H$_2$ conversion
factor for the starburst galaxies.
The non-detections from our new search exclude the previous tentative
($\sim$3$\sigma$) detection of the \co{4}{3} line toward SSA22-LAB01
reported by \citet{Chapman04} with a high significance (12$\sigma$).
We find that the FIR-to-CO luminosity ratio of \lfir/\lpco\ $\gtrsim$
100 \lsun (\unitlco)$^{-1}$ for LABd05 is consistent with the scaling
relations of ULIRG/SMG/QSO/RG.
While our sensitive CO searches already place interesting limits on the
brightest \lya\ blobs, future observations with ALMA will allow us to
routinely detect CO in similar systems (0.1 Jy\,\kms\ for $\sim$1 hour
integration) beyond the sensitivity limits that are accessible today.

%----------------------------------------------------------------------
\acknowledgments

We thank anonymous referee for the prompt report and the helpful comments.
We thank Jan Martin Winters and Melanie Krips for supporting our PdBI
CO and 1.2\,mm observations.  Y.Y.\ thanks Elisabete Da Cunha and Brent
Groves for the helpful discussions on the SED fitting.
The research activities of AD are supported by the National Optical
Astronomy Observatory, which is operated by the Association of
Universities for Research in Astronomy (AURA) under cooperative agreement
with the National Science Foundation.
Based on observations carried out with the IRAM Plateau de Bure
Interferometer and the IRAM-30m Telescope. IRAM is supported by INSU/CNRS
(France), MPG (Germany) and IGN (Spain).
This publication is based on data acquired with the Atacama Pathfinder
Experiment (APEX). APEX is a collaboration between the Max-Planck-Institut
f\"ur Radioastronomie, the European Southern Observatory, and the Onsala
Space Observatory.
This work is based in part on observations made with the Spitzer Space
Telescope, which is operated by the Jet Propulsion Laboratory, California
Institute of Technology under a contract with NASA.
Herschel is an ESA space observatory with science instruments provided
by European-led Principal Investigator consortia and with important
participation from NASA.

\smallskip
\smallskip
Facilities: \facility{PdBI, {\it Spitzer} (MIPS), APEX (LABOCA), 
                      IRAM-30m (MAMBO-2), Herschel (SPIRE)}

\newcommand\ff[1]{\tablenotemark{#1}}
%\begin{deluxetable*}{cccc} % emulateapj
\begin{deluxetable}{cccc}
\tablewidth{0pt}
\tabletypesize{\small}
\tabletypesize{\scriptsize}
%\rotate
\tablecaption{Photometry of LABd05}
\tablehead{
%--------------------
\colhead{Facility}&  
\colhead{$\lambda_{\rm obs}$}&    
\colhead{$S_\nu$ \ff{a}}&
\colhead{Note}\\   
%--------------------
\colhead{}&  
\colhead{(\micron)}&    
\colhead{(mJy)}&
\colhead{}
%--------------------
}
%--------------------
\startdata
% Facility       &  $\lambda_{\rm obs}$ &    $S_\nu$               & % Note                  \\   
%                &            (\micron) &      (mJy)               &                         \\
  Spitzer/MIPS   &                   24 &      0.856  $\pm$  0.005 &  Dey et al.~(2005)      \\
                 &                   70 &      $<$ \bb9            &  this study             \\
                 &                  160 &      $<$ 51              &  this study             \\
\hline\\[-1.5ex] 
 CSO SHARC-II    &                  350 &       37    $\pm$   13   &  Bussmann et al.~(2009) \\
\hline\\[-1.5ex] 
 Herschel SPIRE  &                  250 &      18.8   $\pm$  5.2   &  this study             \\
                 &                  350 &      26.9   $\pm$  5.1   &  this study             \\
\hline\\[-1.5ex] 
 IRAM MAMBO-2    &                 1200 &       2.76  $\pm$  0.35  &  this study                      
\enddata
\tablenotetext{a}{3$\sigma$ upper limits if non-detection.}
\label{tab:photometry}
\end{deluxetable}
%----------------------------------------------------------------------

%----------------------------------------------------------------------
\def\arraystretch{1.0}
\begin{deluxetable}{cccccccc}
\tabletypesize{\small}
\tabletypesize{\scriptsize}
\tablecaption{CO Line Observations for LABd05 and SSA22-LAB01}
\tablehead{
%--------------------
\colhead{Source                  }&  
\colhead{$z_{\rm Ly\alpha}$\ff{a}}&  
\colhead{Transition              }&  
\colhead{$\nu_{\rm obs}$         }&  
\colhead{$\sigma_{\rm rms}$      }&  
\colhead{$\delta v$\ff{b}        }&  
\colhead{\lpco\ ({\sl J}+1$\rightarrow${\sl J})\ff{c}}&  
\colhead{\mhtwo            \ff{d}}\\[0.5ex]
%--------------------
\colhead{                                      }&
\colhead{                                      }&
\colhead{                                      }&
\colhead{(GHz)                                 }&
\colhead{(mJy)                                 }&
\colhead{(\kms)                                }&
\colhead{($10^{10}$ ${\rm K\,km\,s^{-1} pc^2}$)}&
\colhead{($10^{10}$ \msun)                     }
%--------------------
}
%--------------------
\startdata
% Source         &      z             & transition & frequency  &  rms    &  delta v  &                L'CO  &        M(H2) \\
%                &                    &            &            &         &    km/s   &  10^10 K km/s pc^2   &   10^10Msun  \\
  LABd05         & 2.656$\,\pm\,$0.006&     3--2   & \bb94.614  &  2.8\bb &     42.0  &          $<$   3.90  &  $<$   3.12  \\ 
                 &                    &     5--4   &   157.675  &  2.8\bb &     42.0  &          $<$   1.37  &  \bb \nodata \\
\hline\\[-1.5ex]
  SSA22-LAB01    & 3.102$\,\pm\,$0.005&     3--2   & \bb84.299  &  0.53   &     100   &          $<$   1.47  &  $<$   1.17  \\
                 &                    &     4--3   &   112.300  &  2.0\bb &     26.7  &          $<$   1.64  &  \bb \nodata 
\enddata
\label{tab:CO}
\tablenotetext{a}{Redshifts determined from \lya\ lines \citep{Dey05,Bower04}.}
\tablenotetext{b}{Channel bandwidth.}
\tablenotetext{c}{3$\sigma$ upper limits assuming CO linewidth $\Delta V$ = 400\,\kms.}
\tablenotetext{d}{3$\sigma$ upper limits assuming $X_{\rm CO}$ =
0.8 \unitxco\ and constant brightness temperatures.}
\end{deluxetable}
%----------------------------------------------------------------------

%----------------------------------------------------------------------
\begin{deluxetable}{ccccc}
\tablewidth{0pt}
\tabletypesize{\small}
\tabletypesize{\scriptsize}
%\rotate
%\tablecaption{Photometry of SSA22-LAB01}
\tablecaption{FIR Observations of SSA22-LAB01}
\tablehead{
%--------------------
\colhead{Facility}&  
\colhead{Wavelength}&  
\colhead{Beam size}&        
\colhead{$S_\nu$\ff{a}}&
\colhead{Reference\ff{b}} \\
%--------------------
\colhead{}&  
\colhead{(\micron)}&    
\colhead{(arcsec)}&
\colhead{(mJy)}&
\colhead{}
%--------------------
}
%--------------------
\startdata
% Facility  &  Wavelength  &  Beam size     &        Snu              &      \\ %%%             Reference \\
%           &    (micron)  &   (arcsec)     &      (mJy)              &      \\ %%%                       \\
  SCUBA     &     \phn850  &      15        &       16.8 $\pm$ \phn2.9 &  (1) \\ %%% Chapman et al. (2004) \\
            &     \phn450  &       8        &       45.1 $\pm$    15.1 &      \\ %%%                       \\
\hline\\[-1.5ex] 
  AzTEC     &        1100  &      28        & $<$\phn3.0               &  (2) \\ %%%   Kohno et al. (2008) \\
  SMA       &     \phn880  & 2.5$\times$1.9 & $<$\phn4.2               &  (3) \\ %%% Matsuda et al. (2007) \\
\hline\\[-1.5ex] 
  LABOCA    &     \phn870  &      19        & $<$   12.0               &  this study    \\ %%% Matsuda et al. (2007) \\
  PdBI      &        1250  & 2.4$\times$1.6 & $<$   0.45               &  this study    \\ %%%                       \\
  PdBI      &        3500  & 8.4$\times$5.4 & $<$   0.15               &  this study       %%%                       
\enddata
\label{tab:SED_LAB01}
\tablenotetext{a}{3$\sigma$ upper limits if non-detection.}
\tablenotetext{b}{(1) \citet{Chapman01, Chapman04, Chapman05}, 
                  (2) \citet{Kohno08, Tamura09}, 
                  (3) \citet{Matsuda07}}
\end{deluxetable}
%----------------------------------------------------------------------


\begin{thebibliography}{}

\bibitem[Bower et al.(2004)]{Bower04} Bower, R.~G., et al.\ 2004, \mnras, 351, 63 
\bibitem[Bussmann et al.(2009)]{Bussmann09} Bussmann, R.~S., et al.\ 2009, \apj, 705, 184 
\bibitem[Chapman et al.(2001)]{Chapman01} Chapman, S.~C., Lewis, G.~F., Scott, D., Richards, E., Borys, C., Steidel, C.~C., Adelberger, K.~L., \& Shapley, A.~E.\ 2001, \apjl, 548, L17 
\bibitem[Chapman et al.(2004)]{Chapman04} Chapman, S.~C., Scott, D., Windhorst, R.~A., Frayer, D.~T., Borys, C., Lewis, G.~F., \& Ivison, R.~J.\ 2004, \apj, 606, 85 
\bibitem[Chapman et al.(2005)]{Chapman05} Chapman, S.~C., Blain, A.~W., Smail, I., \& Ivison, R.~J.\ 2005, \apj, 622, 772 
\bibitem[Colbert et al.(2006)]{Colbert06} Colbert, J.~W., Teplitz, H., Francis, P., Palunas, P., Williger, G.~M., \& Woodgate, B.\ 2006, \apjl, 637, L89
\bibitem[Colbert et al.(2011)]{Colbert11} Colbert, J.~W., Scarlata, C., Teplitz, H., Francis, P., Palunas, P., Williger, G.~M., \& Woodgate, B.\ 2011, \apj, 728, 59 
\bibitem[Dekel et al.(2009)]{Dekel09} Dekel, A., et al.\ 2009, \nat, 457, 451 
\bibitem[Dey et al.(2005)]{Dey05} Dey, A., et al.\ 2005, \apj, 629, 654 
\bibitem[Dijkstra \& Loeb(2009)]{Dijkstra&Loeb09} Dijkstra, M., \& Loeb, A.\ 2009, \mnras, 400, 1109 
\bibitem[Downes \& Solomon(1998)]{Downes&Solomon98} Downes, D., \& Solomon, P.~M.\ 1998, \apj, 507, 615 
\bibitem[Draine(2003)]{Draine03} Draine, B.~T.\ 2003, \araa, 41, 241 
\bibitem[Dunne et al.(2000)]{Dunne00} Dunne, L., Eales, S., Edmunds, M., et al.\ 2000, \mnras, 315, 115 
\bibitem[Dunne et al.(2003)]{Dunne03} Dunne, L., Eales, S.~A., \& Edmunds, M.~G.\ 2003, \mnras, 341, 589 
\bibitem[Dunne et al.(2011)]{Dunne11} Dunne, L., Gomez, H.~L., da Cunha, E., et al.\ 2011, \mnras, 1395 
\bibitem[Fardal et al.(2001)]{Fardal01} Fardal, M.~A., Katz, N., Gardner, J.~P., Hernquist, L., Weinberg, D.~H., \& Dav{\' e}, R.\ 2001, \apj, 562, 605 
\bibitem[Francis et al.(2001)]{Francis01} Francis, P.~J., et al.\ 2001, \apj, 554, 1001 
\bibitem[Gao \& Solomon(2004)]{Gao&Solomon04} Gao, Y., \& Solomon, P.~M.\ 2004, \apj, 606, 271 
\bibitem[Geach et al.(2005)]{Geach05} Geach, J.~E., et al.\ 2005, \mnras, 363, 1398
\bibitem[Geach et al.(2007)]{Geach07} Geach, J.~E., Smail, I., Chapman, S.~C., Alexander, D.~M., Blain, A.~W., Stott, J.~P., \& Ivison, R.~J.\ 2007, \apjl, 655, L9 
\bibitem[Geach et al.(2009)]{Geach09} Geach, J.~E., et al.\ 2009, \apj, 700, 1 
\bibitem[Goerdt et al.(2010)]{Goerdt10} Goerdt, T., Dekel, A., Sternberg, A., Ceverino, D., Teyssier, R., \& Primack, J.~R.\ 2010, \mnras, 407, 613 
\bibitem[Gordon et al.(2007)]{Gordon07} Gordon, K.~D., et al.\ 2007, \pasp, 119, 1019 
\bibitem[Greve et al.(2005)]{Greve05} Greve, T.~R., et al.\ 2005, \mnras, 359, 1165 
\bibitem[Haiman \& Rees(2001)]{Haiman&Rees01} Haiman, Z., \& Rees, M.~J.\ 2001, \apj, 556, 87 
\bibitem[Haiman, Spaans, \& Quataert(2000)]{Haiman00} Haiman, Z., Spaans, M., \& Quataert, E.\ 2000, \apjl, 537, L5 
\bibitem[Hildebrand(1983)]{Hildebrand83} Hildebrand, R.~H.\ 1983, \qjras, 24, 267 
\bibitem[Jannuzi \& Dey(1999)]{Jannuzi&Dey99} Jannuzi, B.~T., \& Dey, A.\ 1999, Photometric Redshifts and the Detection of High Redshift Galaxies, 191, 111 
\bibitem[Keel et al.(1999)]{Keel99} Keel, W.~C., Cohen, S.~H., Windhorst, R.~A., \& Waddington, I.\ 1999, \aj, 118, 2547  
\bibitem[Kere{\v s} et al.(2005)]{Keres05} Kere{\v s}, D., Katz, N., Weinberg, D.~H., \& Dav{\'e}, R.\ 2005, \mnras, 363, 2
\bibitem[Kere{\v s} et al.(2009)]{Keres09} Kere{\v s}, D., Katz, N., Fardal, M., Dav{\'e}, R., \& Weinberg, D.~H.\ 2009, \mnras, 395, 160
\bibitem[Kohno et al.(2008)]{Kohno08} Kohno, K., et al.\ 2008, Panoramic Views of Galaxy Formation and Evolution, 399, 264 
\bibitem[Matsuda et al.(2004)]{Matsuda04} Matsuda, Y., et al.\ 2004, \aj, 128, 569 
\bibitem[Matsuda et al.(2005)]{Matsuda05} Matsuda, Y., et al.\ 2005, \apjl, 634, L125 
\bibitem[Matsuda et al.(2007)]{Matsuda07} Matsuda, Y., Iono, D., Ohta, K., Yamada, T., Kawabe, R., Hayashino, T., Peck, A.~B., \& Petitpas, G.~R.\ 2007, \apj, 667, 667 
\bibitem[Matsuda et al.(2009)]{Matsuda09} Matsuda, Y., et al.\ 2009, \mnras, 400, L66 
\bibitem[Matsuda et al.(2011)]{Matsuda11} Matsuda, Y., et al.\ 2011, \mnras, 410, L13      
\bibitem[Nilsson et al.(2006)]{Nilsson06} Nilsson, K.~K., Fynbo, J.~P.~U., M{\o}ller, P., Sommer-Larsen, J., \& Ledoux, C.\ 2006, \aap, 452, L23 
\bibitem[Ouchi et al.(2009)]{Ouchi09} Ouchi, M., et al.\ 2009, \apj, 696, 1164
\bibitem[Palunas et al.(2004)]{Palunas04} Palunas, P., Teplitz, H.~I., Francis, P.~J., Williger, G.~M., \& Woodgate, B.~E.\ 2004, \apj, 602, 545 
\bibitem[Pettini et al.(2001)]{Pettini01} Pettini, M., Shapley, A.~E., Steidel, C.~C., Cuby, J.-G., Dickinson, M., Moorwood, A.~F.~M., Adelberger, K.~L., \& Giavalisco, M.\ 2001, \apj, 554, 981 
\bibitem[Polletta et al.(2007)]{Polletta07} Polletta, M., et al.\ 2007, \apj, 663, 81 
\bibitem[Prescott et al.(2008)]{Prescott08} Prescott, M.~K.~M., Kashikawa, N., Dey, A., \& Matsuda, Y.\ 2008, \apjl, 678, L77 
\bibitem[Prescott et al.(2009)]{Prescott09} Prescott, M.~K.~M., Dey, A., \& Jannuzi, B.~T.\ 2009, \apj, 702, 554 
\bibitem[Prescott(2009)]{Prescott09th} Prescott, M.~K.~M.\ 2009, Ph.D.~Thesis.
\bibitem[Riechers et al.(2006)]{Riechers06} Riechers, D.~A., et al.\ 2006, \apj, 650, 604 
\bibitem[Rieke et al.(2009)]{Rieke09} Rieke, G.~H., Alonso-Herrero, A., Weiner, B.~J., P{\'e}rez-Gonz{\'a}lez, P.~G., Blaylock, M., Donley, J.~L., \& Marcillac, D.\ 2009, \apj, 692, 556 
\bibitem[Saito et al.(2006)]{Saito06} Saito, T., Shimasaku, K., Okamura, S., Ouchi, M., Akiyama, M., \& Yoshida, M.\ 2006, \apj, 648, 54 
\bibitem[Savage \& Oliver(2007)]{Savage&Oliver07} Savage, R.~S., \& Oliver, S.\ 2007, \apj, 661, 1339 
\bibitem[Siringo et al.(2009)]{Siringo09} Siringo, G., et al.\ 2009, \aap, 497, 945 
\bibitem[Smith \& Jarvis(2007)]{Smith&Jarvis07} Smith, D.~J.~B., \& Jarvis, M.~J.\ 2007, \mnras, 378, L49 
\bibitem[Smith et al.(2008)]{Smith08} Smith, D.~J.~B., Jarvis, M.~J., Lacy, M., \& Mart{\'{\i}}nez-Sansigre, A.\ 2008, \mnras, 389, 799 
\bibitem[Solomon \& Vanden Bout(2005)]{Solomon05} Solomon, P.~M., \& Vanden Bout, P.~A.\ 2005, \araa, 43, 677 
\bibitem[Steidel et al.(2000)]{Steidel00} Steidel, C.~C., Adelberger, K.~L., Shapley, A.~E., Pettini, M., Dickinson, M., \& Giavalisco, M.\ 2000, \apj, 532, 170 
\bibitem[Steidel et al.(2004)]{Steidel04} Steidel, C.~C., Shapley, A.~E., Pettini, M., Adelberger, K.~L., Erb, D.~K., Reddy, N.~A., \& Hunt, M.~P.\ 2004, \apj, 604, 534 
\bibitem[Steidel et al.(2010)]{Steidel10} Steidel, C.~C., Erb, D.~K., Shapley, A.~E., Pettini, M., Reddy, N., Bogosavljevi{\'c}, M., Rudie, G.~C., \& Rakic, O.\ 2010, \apj, 717, 289 
\bibitem[Tamura et al.(2009)]{Tamura09} Tamura, Y., et al.\ 2009, \nat, 459, 61 
\bibitem[Taniguchi \& Shioya(2000)]{Taniguchi&Shioya00} Taniguchi, Y.~\& Shioya, Y.\ 2000, \apjl, 532, L13 
\bibitem[Uchimoto et al.(2008)]{Uchimoto08} Uchimoto, Y.~K., et al.\ 2008, \pasj, 60, 683 
\bibitem[Weiss et al.(2007)]{Weiss07} Weiss, A., Downes, D., Walter, F., \& Henkel, C.\ 2007, From Z-Machines to ALMA: (Sub)Millimeter Spectroscopy of Galaxies, 375, 25 
\bibitem[Yang et al.(2009)]{Yang09} Yang, Y., Zabludoff, A., Tremonti, C., Eisenstein, D., \& Dav{\'e}, R.\ 2009, \apj, 693, 1579 
\bibitem[Yang et al.(2010)]{Yang10} Yang, Y., Zabludoff, A., Eisenstein, D., \& Dav{\'e}, R.\ 2010, \apj, 719, 1654 
\bibitem[Yang et al.(2011)]{Yang11} Yang, Y., Zabludoff, A., Jahnke, K., Eisenstein, D., Dav{\'e}, R., Shectman, S.~A., \& Kelson, D.~D.\ 2011, \apj, 735, 87 

%\bibitem[Riechers et al.(2011)]{Riechers11} Riechers, D.~A., et al.\ 2011, \apjl, 733, L12 
%\bibitem[Ivison et al.(2011)]{Ivison11} Ivison, R.~J., Papadopoulos, P.~P., Smail, I., Greve, T.~R., Thomson, A.~P., Xilouris, E.~M., \& Chapman, S.~C.\ 2011, \mnras, 412, 1913 
%\bibitem[Blain et al.(2003)]{Blain03} Blain, A.~W., Barnard, V.~E., \& Chapman, S.~C.\ 2003, \mnras, 338, 733 
%\bibitem[Dannerbauer et al.(2009)]{Dannerbauer09} Dannerbauer, H., Daddi, E., Riechers, D.~A., Walter, F., Carilli, C.~L., Dickinson, M., Elbaz, D., \& Morrison, G.~E.\ 2009, \apjl, 698, L178 
%\bibitem[Dowell et al.(2003)]{sharc} Dowell C.D., et al., 2003, SPIE, 4855, 73
%\bibitem[Kov{\'a}cs(2006)]{crush} Kov{\'a}cs, A.\ 2006, Ph.D.~Thesis, Caltech 
%\bibitem[Younger et al.(2009)]{Younger09} Younger, J.~D., et al.\ 2009, \mnras, 394, 1685 
%\bibitem[Zemcov et al.(2003)]{Zemcov03} Zemcov, M., Halpern, M., Borys, C., Chapman, S., Holland, W., Pierpaoli, E., \& Scott, D.\ 2003, \mnras, 346, 1179 
%\bibitem[Wilman et al.(2005)]{Wilman05} Wilman, R.~J., Gerssen, J., Bower, R.~G., Morris, S.~L., Bacon, R., de Zeeuw, P.~T., \& Davies, R.~L.\ 2005, \nat, 436, 227 
%\bibitem[Yang et al.(2006)]{Yang06} Yang, Y., Zabludoff, A.~I., Dav{\'e}, R., Eisenstein, D.~J., Pinto, P.~A., Katz, N., Weinberg, D.~H., \& Barton, E.~J.\ 2006, \apj, 640, 539 

\end{thebibliography}
\end{document}